\documentclass[a4paper,11pt]{article}

\pdfoutput=1 

\usepackage{jheppub} 

\usepackage[x11names,dvipsnames,svgnames]{xcolor}
\usepackage{hyperref}

\usepackage{tikz}
\usetikzlibrary{patterns,arrows,decorations.pathmorphing,decorations.markings}
\usetikzlibrary{decorations.markings}

\usepackage{youngtab}

\usepackage{bbm,slashed,empheq,mathrsfs}
\usepackage[normalem]{ulem}

\usepackage[T1]{fontenc} 

\usepackage{soul}
\usepackage{enumitem} 

\def\rd{{\rm d}}
\def\tr{{\rm Tr}\,}

\def\msbar{$\overline{\text{MS}}$}

\def\nn{ \nonumber \\ }
\def\vev#1{\left\langle #1 \right \rangle}

\def\fcs[#1#2]#3{ f^{\phantom{#1#2}#3}_{#1#2} }

\def\absent#1{\textcolor{red}{ \text{\sout{$#1$} }}}

\newcommand{\cL}{\mathcal{L}}

\preprint{
	\mbox{}\hfill{} ZU-TH 45/23 \\
	\mbox{}\hfill{} PSI-PR-23-29 
}

\title{\boldmath An Algebraic Formula for Two Loop Renormalization of Scalar Quantum Field Theory
}

\author[a]{Elizabeth E.~Jenkins,}

\author[a]{Aneesh V.~Manohar,}

\author[a,b,c]{Luca Naterop,}

\author[a]{Julie Pag\`es}

\affiliation[a]{Physics Department 0319,
University of California San Diego,\\ 9500 Gilman Drive, La Jolla, CA 92093-0319, USA}

\affiliation[b]{Physik-Institut, Universit\"at Z\"urich,
Winterthurerstrasse 190, CH-8057 Z\"urich, Switzerland}

\affiliation[c]{Paul Scherrer Institut CH-5232 Villigen PSI, Switzerland}

\emailAdd{ejenkins@ucsd.edu}
\emailAdd{amanohar@ucsd.edu}
\emailAdd{luca.naterop@physik.uzh.ch}
\emailAdd{jcpages@ucsd.edu}

\abstract{
We find a general formula for the two-loop renormalization counterterms of a scalar quantum field theory with interactions containing up to two derivatives, extending 't~Hooft's one-loop result.  The method can also be used for theories with higher derivative interactions, as long as the terms in the Lagrangian have at most one derivative acting on each field. We show that diagrams with factorizable topologies do not contribute to the renormalization group equations. 
The results in this paper will be combined with the geometric method in a subsequent paper to obtain the counterterms and renormalization group equations for the scalar sector of effective field theories (EFT) to two-loop order.
}

\allowdisplaybreaks

\begin{document} 
\maketitle
\flushbottom

\section{Introduction}\label{sec:intro}

The general one-loop renormalization counterterms for scalar field theory were computed by 't~Hooft~\cite{tHooft:1973bhk} in dimensional regularization in the MS scheme.\footnote{The results are also valid in the \msbar\ scheme, where the parameter $\mu$ of the MS scheme is rescaled, $\mu^2 \to \mu^2 e^\gamma/(4\pi) $, or the usual scheme  in chiral perturbation theory, where  $\mu^2 \to \mu^2 e^{(\gamma-1)}/(4\pi) $. We will refer to all of these as the MS scheme.}  't~Hooft's formula applies to theories with Lagrangians containing up to two derivatives in the fields, and a canonical kinetic energy term.  It allows for the computation of the one-loop renormalization constants and anomalous dimensions by purely algebraic manipulations. The formula was extended to arbitrary two-derivative Lagrangians using a  geometrical method~\cite{Alonso:2015fsp,Alonso:2016oah} which exploits the invariance of the $S$-matrix and physical observables under field redefinitions~\cite{Chisholm:1961tha,Kamefuchi:1961sb,Politzer:1980me,Arzt:1993gz,Manohar:2018aog}.  Field redefinitions can be interpreted as coordinate transformations on the manifold on which the quantum fields live, and the renormalization counterterms can be written in terms of geometric objects such as the Riemann curvature tensor of the scalar manifold $\mathcal{M}$.

The geometrical approach was used in~\cite{Alonso:2015fsp,Alonso:2016oah} to study the scalar sector of chiral perturbation theory, the Standard Model Effective Field Theory (SMEFT), and Higgs Effective Field Theory (HEFT).  It was shown in refs.~\cite{Alonso:2015fsp,Alonso:2016oah} that deviations of longitudinal gauge boson and Higgs cross sections from their SM values measure the Riemann curvature of $\mathcal{M}$. The formalism was generalized to include gauge fields in~\cite{Helset:2022pde,Helset:2022tlf} and fermions in~\cite{Finn:2020nvn,Gattus:2023gep,Assi:2023zid}. The geometric approach reorganizes perturbation theory, and greatly simplifies the computation of radiative corrections. The geometric approach uses an expansion in derivatives rather than operator dimension, so one can include operators of arbitrary high dimension in an EFT with a fixed number of derivatives.  Applications of the method to chiral perturbation theory, which also has a derivative expansion, were given in~\cite{Alonso:2016oah}, and applications to SMEFT were discussed in Ref.~\cite{Buchalla:2019wsc,Helset:2020yio}. Ref.~\cite{Helset:2022pde} computed the one-loop renormalization group equations (RGE) in SMEFT to dimension eight using this method, obtaining some new results as well as cross-checking some recent calculations~\cite{Jenkins:2013zja,Jenkins:2013wua,Alonso:2013hga,Jenkins:2017jig,Jenkins:2017dyc,DasBakshi:2022mwk,DasBakshi:2023htx}. There are also extensions of the method to Lagrangians with higher derivatives~\cite{Cheung:2022vnd,Cohen:2022uuw,Craig:2023hhp,Alminawi:2023qtf,Naterop:qv}, and to higher loops~\cite{Alonso:2022ffe,Alonso:2017tdy}.

In this paper, we compute two-loop counterterms and anomalous dimensions in scalar theories with arbitrary interactions up to two derivatives. We find a general algebraic formula for two-loop counterterms in the MS scheme, generalizing the well-known one-loop result of 't~Hooft~\cite{tHooft:1973bhk}.
We show that a large class of factorizable diagrams do not contribute to the two-loop anomalous dimensions, as they factor into the product of one-loop diagrams. This simplification is hidden in the usual Feynman diagram calculation, which organizes the terms by powers of the coupling constant, rather than topology. We derive the formula for the anomalous dimensions in minimal subtraction and generalize 't~Hooft's consistency relations~\cite{tHooft:1973mfk} for counterterms to an arbitrary EFT for which the renormalization group equations are non-linear. 
The results of this paper are combined with the geometrical methods of~\cite{Alonso:2015fsp,Alonso:2016oah} in a second paper \cite{Naterop:qv} (referred to as paper II), which allows for the computation of two-loop renormalization of an arbitrary EFT with terms up to two derivatives. It also applies to theories with more than two derivatives provided higher order terms contain at most single derivatives acting on each field.  At the end of this paper, we explicitly evaluate the two-loop formula for the renormalizable $O(n)$ model, and verify that it produces the correct anomalous dimensions.  The calculation for the $O(n)$ EFT which includes higher dimension operators is presented in paper II, since it is more efficiently done using geometrical methods.  The second paper uses the  two-loop counterterm formula to compute the two-loop anomalous dimensions for the scalar sector of SMEFT to dimension six, and for chiral perturbation theory to order $p^6$. It also explains how the formalism applies to Lagrangians with higher derivative terms, provided there is only a single derivative acting on each field, as for the $p^4$ Lagrangian in chiral perturbation theory.

\section{Form of the Loop Corrections}\label{sec:form}

\begin{figure}
\begin{center}
\begin{tikzpicture}[scale=0.6]
\draw (0,0) circle (1);
\filldraw (0:1) circle (0.05);
\filldraw (90:1) circle (0.05);
\filldraw (180:1) circle (0.05);
\draw[dashed] (90:1) -- +(60:1.5);
\draw[dashed] (90:1) -- +(90:1.5);
\draw[dashed] (90:1) -- +(120:1.5);
\draw[dashed] (0:1) -- +(20:1.5);
\draw[dashed] (0:1) -- +(-20:1.5);
\draw[dashed] (180:1) -- +(180:1.5);
\end{tikzpicture}
\hspace{2cm}
\begin{tikzpicture}[scale=0.6]
\draw (0,0) circle (1);
\end{tikzpicture}
\end{center}
\caption{One-loop correction to the action. The solid line is the internal field $\eta$, and the dashed lines represent external fields $\overline \phi$. There can be an arbitrary number of vertices, each of which has two $\eta$ lines. The right hand figure shows the corresponding skeleton graph, which can have an arbitrary number of $\eta\eta$ vertex insertions in the loop.
\label{fig:loop}}
\end{figure}
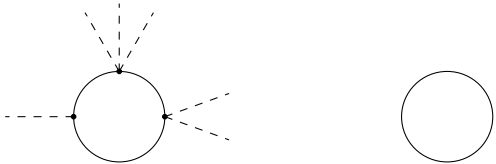

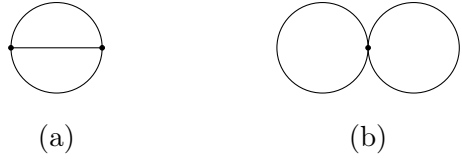
\begin{figure}
\begin{center}
\begin{tikzpicture}[scale=0.6]
\draw (0,0) circle (1);
\draw (-1,0) -- (1,0);
\filldraw (-1,0) circle (0.05);
\filldraw (1,0) circle (0.05);
\draw (0,-2) node {(a)};
\end{tikzpicture}
\hspace{2cm}
\begin{tikzpicture}[scale=0.6]
\draw (-1,0) circle (1);
\draw (1,0) circle (1);
\filldraw (0,0) circle (0.05);
\draw (0,-2) node {(b)};
\end{tikzpicture}
\caption{\label{fig:two} Skeleton graphs for two-loop corrections to the action. There can be an an arbitrary number of $\eta\eta$ vertex insertions, as in Fig.~\ref{fig:loop}.}
\end{center}
\end{figure}

In this section, we compute the general formula for the one-loop and two-loop counterterms in a scalar theory, with terms containing up to two derivatives, in the presence of an arbitrary background field. The one-loop corrections are given by writing the scalar field $\phi \to \overline \phi + \eta$ in terms of a background field $\overline \phi$, and a quantum field $\eta$, which is integrated over and only appears as internal lines in loop graphs. The generic one-loop graph is shown in Fig.~\ref{fig:loop}, and consists of a single $\eta$ loop, with arbitrary insertions of external fields from $\eta^2$ vertices. There are two generic connected one-particle irreducible two-loop graphs, shown in Fig.~\ref{fig:two}, which involve either two insertions of $\eta^3$ vertices, or one insertion of an $\eta^4$ vertex. Both types of graphs can have an arbitrary number of insertions of $\eta^2$ vertices.  These results follow from the topological identity
\begin{align}
(F-2) + 2 L &= \sum_i  (F_i-2)\,,
\label{2.1}
\end{align}
where $F$ is the total number of external fields in a Feynman graph, $F_i$ are the number of fields at each vertex, and $L$ is the number of loops.\footnote{The topological identity was used previously to derive the power counting rule of naive dimensional analysis, which gives a method of counting $4\pi$ factors associated to terms in the Lagrangian. A more detailed discussion can be found in~\cite{Manohar:1983md,Gavela:2016bzc}. The same identity was also used to determine factors of coupling constants associated with a given loop graph~\cite{Jenkins:2013sda}.}
 In our case, there are no external quantum fields, so $F=0$.  Consequently, the left-hand side of Eq.~\eqref{2.1} for $L=1$ and $L=2$ is equal to zero and two, respectively.
Thus, the one-loop graphs only contain quadratic vertices, whereas the two-loop graphs either contain two cubic vertices or one quartic vertex.  
%
%
%
%
%
%
%
%

We now summarize 't~Hooft's calculation~\cite{tHooft:1973bhk}. Starting with a Lagrangian $\cL(\phi)$, we write the scalar field as the sum of a background field and a quantum field $\phi= \overline \phi + \eta$, and expand in $\eta$.\footnote{We focus on scalar fields. The method also works for gauge fields and fermions~\cite{tHooft:1973bhk}.} The general Lagrangian up to order $\eta^2$ and two derivatives is
\begin{align}
{\cal L} &= \frac12 \partial_\mu \eta^T \partial^\mu \eta + \partial_\mu \eta^T N^\mu \eta + \frac12 \eta^T X \eta \,,
\label{3.1a}
\end{align}
where $\eta^a$ are real scalar fields, $N^\mu_{ab}$ and $X_{ab}$ are matrices which are functionals of the background fields, and we use matrix notation for the index contractions. The background fields can be any kind of field, since they do not affect the evaluation of the loop integral over $\eta$. Following~\cite{tHooft:1973bhk}, we assume the kinetic term is canonically normalized. The generalization to an arbitrary kinetic term
\begin{align}
{\cal L} &= \frac12  \partial_\mu \eta^T g\, \partial^\mu \eta 
\label{3.1b}
\end{align}
was given in~\cite{Alonso:2015fsp,Alonso:2016oah}, and will be discussed in paper II. In eq.~\eqref{3.1b}, $g_{ab}$, the metric on the scalar field manifold, is a function of the background fields. Eq.~\eqref{3.1a} corresponds to a trivial scalar metric $g_{ab}=\delta_{ab}$. For this paper, the metric is trivial and we do not distinguish between upper and lower flavor indices.

The indices $a,b,\ldots$ refer to flavor indices. $X_{ab}$ is symmetric in its flavor indices, $X_{ab} = X_{ba}$. 
The Lagrangian can be shifted by a total derivative, which leaves the action unchanged. Consider the shift
\begin{align}
{\cal L} &\to {\cal L} + \partial_\mu \left[ \eta^T Z^\mu \eta \right] = {\cal L} +  \eta^T (  \partial_\mu Z^\mu ) \eta + (\partial_\mu \eta)^T Z^\mu  \eta + \eta^T Z^\mu (\partial_\mu \eta) 
\label{3.2}
\end{align}
where $Z^\mu_{ab}=Z^\mu_{ba}$ is a symmetric matrix. The last term can be rewritten as $\eta^T Z^\mu (\partial_\mu \eta) = (\partial_\mu \eta)^T Z^\mu  \eta $ since $Z^\mu$ is symmetric.
The last two terms lead to a shift in the symmetric part of $N^\mu$, $N^\mu \to N^\mu +  2 Z^\mu $. $Z^\mu$ can be chosen to remove the symmetric part of $N^\mu$, so that $N^\mu$ is antisymmetric, $N^\mu_{ab} = - N^\mu_{ba} $. The term $\partial_\mu Z^\mu$ is symmetric, and can be absorbed into a redefinition of $X$, $X \to X +2 \partial_\mu Z^\mu$. Thus in Eq.~\eqref{3.1a}, we can always require $N_\mu$ to be antisymmetric and $X$ to be symmetric.

Define the covariant derivative
\begin{align}
D_\mu \eta &\equiv \partial_\mu \eta + N_\mu \eta \,.
\label{3.3}
\end{align}
$N^\mu$ plays the role of a background $O(n)$ gauge field, since it is a real antisymmetric matrix. The Lagrangian Eq.~\eqref{3.1a} is
\begin{align}
{\cal L} &= \frac12 (D_\mu \eta)^T (D^\mu \eta)+ \frac12 \eta^T (X-N_\mu^T N^\mu) \eta =  \frac12 (D_\mu \eta)^T (D^\mu \eta)+ \frac12 \eta^T (X+N_\mu N^\mu) \eta
\label{3.4}
\end{align}
which is equivalent to
\begin{align}
{\cal L} &= \frac12 (D_\mu \eta)^T (D^\mu \eta)+ \frac12 \eta^T  X \eta
\label{3.4a}
\end{align}
after further redefining $X$ by $X \to X + N_\mu N^\mu$. As noted by  't~Hooft, Eq.~\eqref{3.4a} has an $O(n)$ symmetry,
\begin{align}
\eta & \to O \eta \,, & X &\to O X O^T \,,  & N_\mu & \to O N_\mu O^T - \partial_\mu O O^T \,.
\label{3.4b}
\end{align}
under which $N_\mu$ transforms as a gauge field.
This $O(n)$ symmetry greatly simplifies the computation of the counterterm Lagrangian. Note that $O(n)$ need not be a symmetry of the original Lagrangian. The above method writes the fluctuation Lagrangian in an $O(n)$ symmetric form, even if the starting Lagrangian is not $O(n)$ symmetric.

't~Hooft computed the one-loop counterterm Lagrangian 
\begin{align}
{\cal L}_{\text{c.t.}}^{(1)}  &=   \frac{1}{16 \pi^2 \epsilon} \left[-\frac14 \tr X^2 - \frac{1}{24} \tr Y_{\mu \nu} Y^{\mu \nu} \right]
\label{eq:oneloopct}
\end{align}
in $d=4-2\epsilon$ dimensions in the MS scheme, where
\begin{align}
Y_{\mu \nu} &= [D_\mu,D_\nu] = \partial_\mu N_\nu - \partial_\nu N_\mu + [N_\mu,N_\nu]\,
\label{3.6}
\end{align}
is the $O(n)$ field-strength tensor constructed from $N_\mu$. $Y_{\mu\nu}$ is antisymmetric in its Lorentz indices $Y^{\mu\nu}_{ab}=-Y^{\nu\mu}_{ab}$, as well as in its flavor indices, $Y^{\mu\nu}_{ab}=-Y^{\mu\nu}_{ba}$. The 1-loop counterterm Lagrangian Eq.~\eqref{eq:oneloopct} is gauge-invariant under the $O(n)$ symmetry of Eq.~\eqref{3.4a}.


The two-loop corrections require terms cubic and quartic in $\eta$. Up to two derivatives, the allowed terms are\footnote{ Note that the original Lagrangian need not be $O(n)$ symmetric, but the fluctuations can still be written in $O(n)$ form. The vertical bar $|$ separates the indices contracted with $D_\mu \eta$ from those contracted with $\eta$.} 
\begin{align}
{\cal L} &= A_{abc} \eta^a \eta^b \eta^c + A^\mu_{a|bc} (D_\mu \eta)^a \eta^b \eta^c + A^{\mu \nu} _{ab|c} (D_\mu \eta)^a (D_\nu \eta)^b \eta^c \nn
& + B_{abcd} \eta^a \eta^b \eta^c \eta^d + B^\mu_{a|bcd} (D_\mu \eta)^a \eta^b \eta^c \eta^d+ B^{\mu \nu} _{ab|cd} (D_\mu \eta)^a (D_\nu \eta)^b \eta^c \eta^d \,.
\label{3.7}
\end{align}
The terms are written in terms of covariant derivatives rather than ordinary derivatives, to make use of the $O(n)$ symmetry of the quadratic Lagrangian eq.~\eqref{3.4a}. This can always be done by the replacement $\partial_\mu \to D_\mu - N_\mu$, and absorbing the $N_\mu$ terms into redefinitions of the lower derivative terms in eq.~\eqref{3.7}. The $A$ and $B$ coefficients are arbitrary functionals of the background fields. Terms with $D^2\eta$ can be converted to the form Eq.~\eqref{3.7} by integration by parts.
$A_{abc}$ is completely symmetric in $abc$, $B_{abcd}$ is completely symmetric in $abcd$,
$A^\mu_{a|bc}$ is symmetric in $bc$, $B^\mu_{a|bcd}$ is completely symmetric in $bcd$. The other coefficients satisfy the symmetry relations $A^{\mu \nu}_{ab|c} = A^{\nu \mu}_{ba|c}$, $B^{\mu \nu}_{ab|cd} = B^{\nu \mu}_{ba|cd}  = B^{\mu \nu}_{ab | dc} $.Thus, the coefficients are completely symmetric in flavor indices on each side of the vertical bar, with Lorentz indices coupled to the flavor indices to the left of the bar.  We will always assume these symmetries on the coefficients. As in Eq.~\eqref{3.2}, we can add total derivatives
\begin{align}
{\cal L} &\to {\cal L} + D_\mu \left[ C^\mu_{abc} \eta^a \eta^b \eta^c \right] + D_\mu \left[ F^\mu_{abcd} \eta^a \eta^b \eta^c \eta^d \right] 
\label{3.8}
\end{align}
where $C^\mu_{abc}$ and $D^\mu_{abcd}$ are completely symmetric, which can be used to eliminate the totally symmetric pieces in $A^{\mu}_{a|bc}$ and $B^{\mu}_{a|bcd}$ while shifting $A_{abc}$ and $B_{abcd}$. Thus one can eliminate the totally symmetric parts of $A^{\mu}_{abc}$ and $B^{\mu}_{a|bcd}$,
\begin{align}
A^{\mu}_{a|bc} + A^{\mu}_{b|ca} + A^{\mu}_{c|ab} &= 0\,, \nn
B^{\mu}_{a|bcd} + B^{\mu}_{b|cda} + B^{\mu}_{c|dab} + B^{\mu}_{d|abc} &= 0 \,.
\label{3.9}
\end{align}
$A^\mu_{a|bc}$ has three flavor indices, and so can be decomposed into irreducible representations of the symmetric group on three elements, $S_3$. Similarly $B^\mu_{a|bcd}$ is decomposed into irreducible representations of $S_4$. The irreducible representations of $S_n$ are denoted by Young tableaux with $n$ boxes. The coefficients $A^{\mu}_{a|bc}$ and $B^{\mu}_{a|bcd}$ transform as the irreducible representations with Young tableaux
\Yvcentermath1
\begin{align}
A^\mu & \sim \yng(2,1) &
B^\mu & \sim \yng(3,1)
\label{3.10}
\end{align}
under the symmetric group because of the constraints eq.~\eqref{3.9}. We cannot simplify $A^{\mu \nu}$ and $B^{\mu \nu}$ by adding total derivatives as in Eq.~\eqref{3.8} without introducing terms with two derivatives on a single $\eta$. We will assume that the coefficients have been put in a standard form so that Eqs.~\eqref{3.9} are satisfied.

The two-loop counterterms, from the graphs in Fig.~\ref{fig:two}, either have one quartic $B$-vertex, or two cubic $A$-vertices. The possible two-loop counterterms can be determined by dimensional analysis. The mass dimensions of the coefficients in Eqs.~\eqref{3.4a}, \eqref{3.6} and \eqref{3.7} are listed in Table~\ref{tab:massdimensions}.
%
%
\begin{table}
	\centering
	\begin{tabular}{cccccccccc}
	      & $D_\mu$ & $X$ & $Y^{\mu\nu}$ & $A$ & $A^\mu$ & $A^{\mu\nu}$ & $B$ & $B^\mu$ & $B^{\mu\nu}$ \\ \hline
	Mass dimension & 1       & 2   & 2            & 1   & 0       & $ -1   $        & 0   &  $-1 $     & $-2    $      
	\end{tabular}
	\caption{Mass dimensions of the building blocks.}
	\label{tab:massdimensions}
\end{table}
%
%
Since the theory has $O(n)$ symmetry, the allowed factors from the quadratic Lagrangian are $X$ and $Y_{\mu \nu}$. The possible structures for the two-loop counterterms are given in Table~\ref{tab:ctAstructures} and Table~\ref{tab:ctBstructures}.
%
%
\begin{table}
	\centering
	\begin{align}
	\begin{array}{c | l}
	A A   &  D^2,\ X ,\ \absent{Y}            \\ 
	A^\mu A & \absent{D^3},\ X D ,\ Y D               \\  
	A^\mu A^\mu  & D^4,\ X D^2 ,\ Y D^2 ,\ X^2,\ XY,\ Y^2   \\ 
	A^{\mu\nu} A  & D^4,\ X D^2 ,\ Y D^2 ,\ X^2,\ XY,\ Y^2   \\ 
	A^{\mu\nu} A^\mu &  D^5,\ X D^3 ,\ Y D^3 ,\ X^2 D ,\ XY D,\ Y^2 D   \\ 
	A^{\mu\nu} A^{\mu\nu}  & D^6,\ X D^4,\ Y D^4,\ X^2 D^2,\ XY D^2,\ Y^2 D^2,\ X^3,\ X^2 Y,\ XY^2,\ Y^3                                                            
	\end{array}
	\end{align}
	\caption{Possible two-loop counterterms from Fig.~\ref{fig:two}(a). The first column gives the two cubic vertices in the diagram. Each line gives the possible counterterms    	 for those cubic vertices. For instance,
	the first line means that allowed counterterms with two factors of $A$ can either have two factors of $A$ and two derivatives, or two factors of $A$ and one $X$, with all possible Lorentz and flavor contractions. $\absent{Y}$ means there is no  $A A Y$ counterterm allowed by the Lorentz and flavor contractions.
	The counterterms with $A^{\mu \nu}$ are not needed for our results. The explicit form for the counterterms is given in Sec.~\ref{sec:two-loop-ct}. }
	\label{tab:ctAstructures}
	\end{table}
%
%
%
%
%
\begin{table}
	\centering
	\begin{align}
	\begin{array}{c | l}
	B   &  \absent{D^4},\ \absent{X D^2},\ \absent{Y D^2},\ X^2  , \absent{XY},\ \absent{Y^2}            \\ 
	B^\mu & \absent{D^5},\ \absent{X D^3},\ \absent{Y D^3}\ X^2 D ,\ XY D ,\ \absent{Y^2 D}             \\  
	B^{\mu\nu}  & \absent{D^6},\ X^2 D^2,\ XY D^2,\ Y^2 D^2,\ X^3,\ X^2 Y,\ XY^2,\ Y^3                                                            
	\end{array}
	\end{align}
	\caption{Possible two-loop counterterms from Fig.~\ref{fig:two}(b). The first column gives the single quartic vertex in the diagram.  Each line gives the possible counterterms  for that quartic vertex.  Some counterterms with a line through them are not allowed because they are total derivatives, or because  the diagram factors into a product of one-loop graphs (see text). The explicit counterterms are given in Sec.~\ref{sec:two-loop-ct}. }
	\label{tab:ctBstructures}
	\end{table}
%
%
Some possible counterterms permitted by dimensional analysis are not allowed, and have a line through them. For example, $AAY$ is removed because the Lorentz index contraction implies it must contain $Y^\alpha{}_\alpha=0$. The $AA^\mu D^3$ term vanishes because $A^\mu$ has no completely symmetric flavor piece, from Eq.~\eqref{3.9}.  The $B D^4$ counterterm vanishes, since it is a total derivative. Terms such as $BXD^2$  are not allowed because the $B$-graphs factor into the product of two one-loop graphs, one of which is a scaleless power divergent integral which vanishes. The factorization property of $B$-graphs leads to important consequences, and is discussed in Sec.~\ref{sec:factorizable}.

There is one major simplification that we make at this point. We will see in paper II that the cubic variation of the action does not generate the two-derivative  term $A^{\mu \nu}_{abc} (D_\mu \eta)^a (D_\nu \eta)^b \eta^c$ if one uses Riemann normal coordinates, which were used in~\cite{Alonso:2016oah} to simplify the one-loop calculation. We will therefore drop $A^{\mu \nu}$, which gets rid of many terms in Table~\ref{tab:ctAstructures}. A naive expansion using $\phi \to \overline \phi + \eta$ \emph{does} generate $A^{\mu \nu}$. Dropping $A^{\mu \nu}$ greatly simplifies the computation of the counterterm Lagrangian, and shows the advantages of a geometrical approach. In a renormalizable theory, $A^\mu_{a|bc}$, $B^\mu_{a|bcd}$ and $B^{\mu \nu}_{ab|cd}$ vanish. In paper II, we will see that $A^\mu_{a|bc}$ and  $B^{\mu \nu}_{ab|cd}$ start at dimension six, and  $B^\mu_{a|bcd}$ at dimension eight.

\section{One-Loop Subdivergences}\label{sec:oneloop}

%
%
\begin{table}
	\centering
	\begin{align}
	\begin{array}{c | l}
	A A \eta \eta   &  \mathbbm{1}            \\ 
	A^\mu A  \eta \eta & D               \\  
	A^\mu A^\mu   \eta \eta & D^2,\ X ,\ Y    \\ 
	B \eta \eta & \absent{D^2},\ X,\ \absent{Y} \\
	B^\mu \eta \eta & \absent{D^3},\ XD,\ YD \\                                                   
	B^{\mu \nu} \eta \eta &  \absent{D^4},\ X D^2 ,\ Y D^2 ,\ X^2,\ \absent{XY},\ Y^2     \\                                          
	\end{array}
	\end{align}
	\caption{Possible one-loop subdivergence counterterms from Fig.~\ref{fig:sub}.  The first column gives the two cubic vertices or the single quartic vertex in the one-loop diagram, as well as the two external scalar fields $\eta$.  Each line gives the possible counterterms for the given vertices and fields.   Some counterterms with a line through them are not allowed because they are total derivatives or vanish. The explicit counterterms are given in Sec.~\ref{sec:oneloop}.}
	\label{tab:sub}
	\end{table}
%
%
%

\begin{figure}
\begin{center}
\begin{tikzpicture}[scale=0.6]
\draw (0,0) circle (1);
\filldraw (0:1) circle (0.05);
\filldraw (180:1) circle (0.05);
\draw (1,0) -- (2.25,0);
\draw (-1,0) -- (-2.25,0);
\draw (0,-2.5) node {(a)};
\end{tikzpicture}
\hspace{2cm}
\begin{tikzpicture}[scale=0.6]
\draw (0,0) circle (1);
\draw (-90:1) circle (0.05);
\draw (-90:1) -- +(-45:1.25);
\draw (-90:1) -- +(-135:1.25);
\draw (0,-2.5) node {(b)};
\end{tikzpicture}
\end{center}
\caption{One-loop subdivergence graphs from the $\eta^3$ and $\eta^4$ interactions. The external lines are $\eta$ fields.
\label{fig:sub}}
\end{figure}
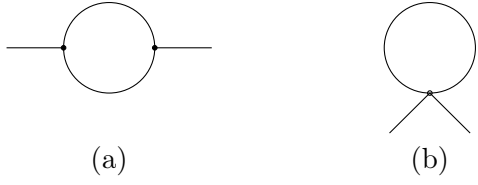

The two-loop graphs in Fig.~\ref{fig:two} contain subdivergences, which have to be subtracted by the insertions of the one-loop counterterms vertices into one-loop graphs. These are generated by the $A$ and $B$ vertices, and have two external $\eta$ fields. The  expressions for the one-loop counterterms allowed by dimensional analysis are listed in Table~\ref{tab:sub}.
They can be computed from the one-loop graphs in Fig.~\ref{fig:sub} using the Lagrangian Eq.~\eqref{3.7} and Eq.~\eqref{3.4a} for the interaction vertices.
%
%
%
%

There is an alternate algebraic way of computing the one-loop subdivergence counterterms which is instructive. Treat Eq.~\eqref{3.4a} plus Eq.~\eqref{3.7} as the new Lagrangian, and repeat the procedure used to obtain Eq.~\eqref{eq:oneloopct}: 
\begin{itemize}
\item[(a)] Make a shift $\eta \to \overline \eta + \zeta$ of the quantum field $\eta$ into a new background quantum field $\overline \eta$ and quantum quantum field $\zeta$. 
\item[(b)] Since we want the one-loop correction,  expand the Lagrangian to quadratic order in $\zeta$. 
\end{itemize}
Shifting the field $\eta \to \overline \eta + \zeta $, expanding to $\mathcal{O}(\zeta^2)$, and dropping the bar over $\eta$, yields
\begin{align}
\cL  &= \frac12 (D_\mu \zeta)^a (D^\mu \zeta)^a+ \frac12 X_{ab} \zeta^a \zeta^b + 3 A_{abc} \zeta^a \zeta^b \eta^c + 2 A^\mu_{a|bc} (D_\mu \zeta)^a \zeta^b \eta^c
+ A^\mu_{a|bc} (D_\mu \eta)^a \zeta^b \zeta^c  \nn
& + 6 B_{abcd} \zeta^a \zeta^b\eta^c \eta^d + 3 B^\mu_{a|bcd} (D_\mu \zeta)^a \zeta^b \eta^c \eta^d
+  3 B^\mu_{a|bcd} (D_\mu \eta)^a \zeta^b \zeta^c \eta^d \nn
&+  B^{\mu \nu}_{ab|cd} (D_\mu \eta)^a (D_\nu \eta)^b \zeta^c \zeta^d
+  4 B^{\mu \nu}_{ab|cd} (D_\mu \eta)^a (D_\nu \zeta)^b \zeta^c \eta^d +  B^{\mu \nu}_{ab|cd} (D_\mu \zeta)^a (D_\nu \zeta)^b \eta^c \eta^d \ .
\label{3.11}
\end{align}
The last term gives a non-trivial  spacetime metric for $\zeta$
\begin{align}
\eta^{\mu \nu} g_{ab} & \to \eta^{\mu\nu} \delta_{ab} +  B^{\mu \nu}_{ab|cd} \eta^c \eta^d
\label{3.12}
\end{align}
on comparing with Eq.~\eqref{3.1a}. This piece cannot be handled by 't~Hooft's formula Eq.~\eqref{eq:oneloopct} so it is treated separately. It can be included if we compute the diagrams explicitly, and is included in the final expressions for the one-loop subdivergence counterterms. 

\begin{itemize}

\item[(c)] Remove the symmetric part of the $\partial \zeta_a \zeta_b$ term by adding a total derivative as in Eq.~\eqref{3.2}. 

\item[(d)] Determine the covariant derivative $D$, and mass term $X$ for the resultant Lagrangian, analogous to the terms in the original Lagrangian Eq.~\eqref{3.4a}. 

\item[(e)] Finally, compute the one-loop counterterms with two external $\bar \eta$ fields from Eq.~\eqref{eq:oneloopct}. This procedure uses a double shift of the field $\phi$ to determine the counterterms, first by $\eta$ and then by $\zeta$. The RGE commute with shifts in the field, i.e.\ computing the RGE and shifting the field gives the same result as shifting the field and then computing the RGE~\cite{Manohar:2020nzp}.

\end{itemize}

The symmetric part  of $(D\zeta) \zeta$ is moved to $X$ by integration by parts, using the identity as in Eq.~\eqref{3.2},
\begin{align}
(D_\mu \zeta)^a \zeta^b + \zeta^a (D_\mu \zeta)^b =  D_\mu(\zeta^a\zeta^b)\,.
\label{3.13}
\end{align}
With this transformation, the Lagrangian Eq.~\eqref{3.11} has the form
\begin{align}
\cL  &= \frac12 (D_\mu \zeta)^T (D^\mu \zeta)  + (D_\mu \zeta)^T U^\mu \zeta + \frac12 \zeta^T \widetilde X \zeta \nn
&= \frac12 (\partial_\mu  \zeta +N_\mu  \zeta + U_\mu \zeta ) ^T (\partial^\mu  \zeta +N^\mu  \zeta + U^\mu \zeta )  + \frac12 \zeta^T \left(\widetilde X
+  U_\mu U^\mu \right) \zeta  
\label{3.14}
\end{align}
where
\begin{align}
U^\mu_{ab} &= (A^\mu_{a|bc} - A^\mu_{b|ac}) \eta^c  + \frac32 (B^\mu_{a|bcd}-B^\mu_{b|acd} )  \eta^c \eta^d
+   2 (B^{\nu \mu}_{c a | bd } - B^{\nu \mu}_{c b|  ad} ) (D_\nu \eta)^c \eta^d
\label{3.15}
\end{align}
and
\begin{align}
\widetilde X_{ab} & = X_{ab} + 6 A_{abc} \eta^c + 2 A^\mu_{c|ab} (D_\mu \eta)^c + 12 B_{abcd} \eta^c \eta^d
+  6 B^\mu_{c|abd}(D_\mu \eta)^c \eta^d   + 2 B^{\mu \nu}_{cd|ab} (D_\mu \eta)^c (D_\mu \eta)^d \nn
&  - D_\mu [ (A^\mu_{a|bc} + A^\mu_{b|ac}) \eta^c]- \frac32 D_\mu[ (B^\mu_{a|bcd} + B^\mu_{b|acd} )  \eta^c \eta^d  ] 
-2 D_\nu[ (B^{\mu \nu}_{ca|bd} + B^{\mu \nu}_{cb|ad} ) (D_\mu \eta)^c \eta^d ]\ .
\label{3.16}
\end{align}
The new covariant derivative and mass term are given by the replacement in Eq.~\eqref{3.4a}
\begin{align}
D_\mu & \to D_\mu + U_\mu = \partial_\mu   +  N_\mu + U_\mu   , \nn
X & \to \widetilde X+  U_\mu U^\mu,
\label{3.17}
\end{align}
with the new field-strength
\begin{align}
Y_{\mu \nu} & \to Y_{\mu \nu} + D_\mu U_\nu - D_\nu U_\mu + [U_\mu,U_\nu] \,,
\label{3.18}
\end{align}
where $D_\mu$ is the old covariant derivative (without $U_\mu$).
Using 't~Hooft's formula Eq.~\eqref{eq:oneloopct}, and retaining the terms quadratic in $A$ or linear in $B$ gives the results listed below in Eq.~\eqref{3.19} and Eq.~\eqref{3.20}.

\subsection{One-Loop Subdivergence Counterterms}\label{sec:one-loop-ct}

The $B$ counterterms are computed from the graph \ref{fig:sub}(b) from the $\eta^4$ terms in eq.~\eqref{3.7}. 
 Since the two-loop graph factorizes into the product of one-loop graphs, it is convenient to first compute one-loop counterterms from a single insertion of the $\eta^2$ interaction Lagrangian
\begin{align}
\mathcal{L}_\text{int} &= C_{ab} \eta^a \eta^b + C^\mu_{ab} (D_\mu \eta)^a \eta^b + C^{\mu \nu}_{ab} (D_\mu \eta)^a (D_\nu \eta)^b
\end{align}
treated as a perturbation added to the Lagrangian eq.~\eqref{3.4a}. This computes the expectation value of $\eta$ bilinears $\vev{\eta^a \eta^b}$, $\vev{D_\mu \eta^a \eta^b }$ and $\vev{D_\mu \eta^a D_\nu \eta^b}$ in the presence of $X$ and $N_\mu$ fields and gives the counterterms
\begin{align}
\mathcal{L}_{\text{c.t.}}  &= \frac{1}{16 \pi^2 \epsilon} \biggl\{ -C_{ab} X_{ba} -\frac12 C^\mu_{ab} D_\mu X_{ba} + \frac16 C^\mu_{ab} D_\alpha Y^{\alpha \mu}_{ba} \nn
& - \frac{1}{12} C^{\mu \mu}_{ab} D^2 X_{ba}  - \frac{1}{12}  C^{\mu \nu}_{ab} \{ D_\mu , D_\nu\} X_{ba} -  \frac{1}{12}  C^{\mu \nu}_{ab} D^2 Y^{\mu \nu}_{ba} 
+  \frac14 C^{\mu \mu}_{ab} X_{bc} X_{ca}  \nn
& + \frac14 C^{\mu \mu}_{ab} (Y^{\mu \nu}_{bc} X_{ca} +  X_{bc} Y^{\mu \nu}_{ca} ) + \frac1{12}  C^{\mu \nu}_{ab} Y^{\nu \alpha}_{bc} Y^{\mu \alpha}_{ca} 
- \frac1{4}  C^{\mu \nu}_{ab} Y^{\mu \alpha}_{bc} Y^{\nu \alpha}_{ca} + \frac1{24}  C^{\mu \mu}_{ab} Y^{\alpha \beta}_{bc} Y^{\alpha \beta}_{ca}
\label{3.11a}
\biggr\}
\end{align} 
\newpage
\noindent
The allowed terms are (schematically) $\vev{\eta_a \eta_b} \to X_{ab}$, $\vev{D_\mu \eta_a \eta_b} \to D_\mu X_{ab} ,\ D_\alpha Y^{\alpha \mu}_{ab}$, \\
$\vev{D_\mu \eta_a D_\nu \eta_b} \sim  D^2 X,  D^2 Y, X^2, XY, Y^2$. $\vev{\eta \eta}$ cannot produce a $Y_{\mu \nu}$, because the Lorentz indices would have to be contracted, and $Y_{\alpha \alpha}=0$. This explains some of the missing entries in Table~\ref{tab:ctBstructures}. Eq.~\eqref{3.11} will be needed for some results in paper II.

The one-loop $B$ subdivergence counterterms are
\begin{align}
\cL_{\text{c.t.}}^{(B,1)} &= \frac{1}{16 \pi^2 \epsilon} \Biggl[ - 6 B_{abcd} X_{cd}  \eta^a \eta^b  - 3 B^\mu_{a | bcd} X_{cd}  (D_\mu\eta)^a \eta^b  - B^{\mu \nu}_{ab | cd} X_{cd}  ( D_\mu\eta)^a
(D_\nu \eta)^b  \nn
& -\frac32 B^\alpha_{a | bcd} (D_\alpha X)_{ab}  \eta^c \eta^d   -2 B^{\alpha \nu}_{a c|  b d}  (D_\alpha X)_{ab}  (D_\nu \eta)^c \eta^d   \nn
&  + \frac14 (D_\mu Y_{\mu \alpha})_{ba}  \left( B^{\alpha }_{a | bcd}-B^{\alpha }_{b | acd} \right) \eta^c \eta^d + \frac13 (D_\mu Y_{\mu \alpha})_{ba}  \left( B^{\alpha \nu}_{a c | bd}-B^{\alpha \nu}_{b c| ad} \right) (D_\nu \eta)^c \eta^d  \nn
& \biggl[ -\frac{1}{12} (D^2 X)_{ab}  B^{\alpha \alpha}_{ab | cd }  -\frac{1}{12}  ( \{D_\mu, D_\nu \} X )_{ab} B^{\mu \nu}_{ab | cd}
-  \frac{1}{12} B^{\mu \nu}_{ab | cd } (D^2 Y^{\mu \nu})_{ba}  \nn
& +  \frac{1}{4} B^{\alpha \alpha}_{ab | cd } X_{ae} X_{eb}  + \frac{1}{12}  B^{\mu \nu}_{ab | cd } Y^{\nu \alpha}_{be}  Y^{\mu \alpha}_{ea} - \frac14   B^{\mu \nu}_{ab | cd } Y^{\mu \alpha}_{be} Y^{\nu \alpha}_{ea} +  \frac{1}{24}     B^{\mu \mu}_{ab| cd } Y^{\alpha \beta}_{be} Y^{\alpha \beta}_{ea}  \biggr]  \eta^c \eta^d \Biggr] \ .
\label{3.19}
\end{align}
These can be obtained by starting with the $B$ Lagrangians eq.~\eqref{3.4a},  choosing two $\eta$ fields to be in the loop, and using eq.~\eqref{3.11a} for the result of the loop graph. The coefficients in eq.~\eqref{3.19} differ from those in eq.~\eqref{3.11a} by the combinatorial factor for picking two $\eta$ fields out of a $\eta^4$ vertex.
The results in the first three rows were also obtained by applying 't~Hooft's method, which gives the same result.
The results in the last two rows could not be computed using  't~Hooft's method  because they arise from terms in Eq.~\eqref{3.12} giving a non-trivial spacetime metric.

The one-loop subdivergence counterterms from the $A$-type terms were computed from the one-loop graphs, and by 't~Hooft's method, both of which give the result
\begin{align}
\cL_{\text{c.t.}}^{(A,1)} &= \frac{1}{(16 \pi^2)} \frac{1}{\epsilon} \Biggl[-9 A_{c a b } A_{ d a b } \eta^c \eta^d  + 6 A_{ a b c }   (D_\mu  A^\mu_{a | b d }  )\eta^c \eta^d + 18  A_{ a b c }    A^\mu_{a | b d }\eta^c (D_\mu \eta)^d  \nn
& + 2 D_\mu \left( A^\mu_{a | b d } \eta^d \right)( A^\nu_{d | a b } (D_\nu \eta)^d - A^\mu_{c | a b } A^\nu_{d | a b } (D_\mu \eta)^c (D_\nu \eta)^d  +
 \frac16 D_\alpha \left( A^\mu_{a | b c }  \eta^c \right)  D_\alpha \left( A^\mu_{a | b d } \eta^d \right)  \nn
 &  - \frac2 3 D_\mu \left( A^\mu_{a | b c }  \eta^c \right)  D_\nu \left( A^\nu_{a | b d } \eta^d \right) +  \frac12 \left( A^\mu_{a | b c }  \eta^c \right) X_{af} \left( A^\mu_{f | b d } \eta^d \right) 
 + \frac12 \left( A^\mu_{a | b c }  \eta^c \right) X_{ b f } \left( A^\mu_{a | f d } \eta^d \right) \nn
 & - \frac16 D_\alpha \left( A^\mu_{a | b c }  \eta^c \right)  D_\alpha \left( A^\mu_{b | a d } \eta^d \right)   - \frac1 3 D_\mu \left( A^\mu_{a | b c }  \eta^c \right)  D_\nu \left( A^\nu_{b | a d } \eta^d \right)   -  \left( A^\mu_{a | b c }  \eta^c \right) X_{a f } \left( A^\mu_{b | f d } \eta^d \right) \nn
 &+  \frac23 Y^{\mu \nu}_{ a b }  A^\nu_{b | e c}  A^\mu_{e | a d}  \eta^c \eta^d  -  \frac13 Y^{\mu \nu}_{ a b }  A^\mu_{a | e c}  A^\nu_{b | e d} \eta^c \eta^d- \frac13 Y^{\mu \nu}_{a b } A^\mu_{e | a c}  A^\nu_{e | b d } \eta^c \eta^d \Biggr] \ .
\label{3.20}
\end{align}

The one-loop counterterms with quantum fluctuations Eq.~\eqref{3.19} and Eq.~\eqref{3.20} are auxiliary results needed for the computation of two loop counterterms to the original Lagrangian. They are not included in the counterterms for the original theory. The one-loop counterterms for the original theory were already given in Eq.~\eqref{eq:oneloopct}. The results Eq.~\eqref{3.19} and Eq.~\eqref{3.20} can be obtained from Eq.~\eqref{eq:oneloopct} by making the replacement $\phi \to \phi + \eta$ and expanding to quadratic order in the quantum fields.

\section{Two-Loop Counterterms}
\label{sec:two-loop-ct}

The two-loop counterterms are computed from the graphs in Fig.~\ref{fig:two} and the one-loop graphs with one-loop counterterm vertices computed in the previous section. The resulting two-loop divergences are local, with the non-local pieces all cancelling between the two-loop graphs and the one-loop subdivergence graphs. This cancellation provides a non-trivial check on the computation. The details of the calculational method are given in Appendix~\ref{appendix:uvextraction}.

\subsection{Quartic $B$ terms}\label{sec:quartic}

The $B$-type two-loop counterterms from the figure eight topology Fig.~\ref{fig:two}(b) are
\begin{align}
\cL^{(B,2)}_{\text{c.t.}} & =  \frac{1}{(16 \pi^2)^2 \epsilon^2 } \Biggl[ 3 B_{a b c d } X_{a b } X_{c d } +  \frac{3}{2} B^\alpha_{a | b c d } (D_\alpha X)_{a b } X_{c d }
+  \frac{1}{2}  B^{\alpha }_{a | b c d }  (D_\mu Y_{\mu \alpha})_{ab } X_{c d } \nn
& +  \frac{1}{12}   B^{\alpha \alpha}_{a b | c d } (D^2 X)_{a b }  X_{c d }  +   \frac{1}{12}  B^{\mu \nu}_{a b | c d }   ( \{D_\mu, D_\nu \} X)_{a b }X_{c d }
 + \frac{1}{12 } B^{\mu \nu}_{a b | c d } (D^2 Y^{\mu \nu})_{a b  }  X_{c d }  \nn
 & - \frac{1}{4}  B^{\alpha \alpha}_{a b | c d } X_{a e } X_{e b } X_{c d } + \frac1{4 }  B^{\mu \nu}_{a b | c d } ( X_{ae  } Y^{\mu \nu}_{eb } +  Y^{\mu \nu}_{a e} X_{e b } )X_{c d } \nn
 &   -\frac{1}{12} B^{\mu \nu}_{a b | c d }  Y^{\mu \alpha}_{ae } Y^{\nu \alpha}_{e b } X_{c d } +\frac14  B^{\mu \nu}_{a b | c d }Y^{\nu \alpha}_{a e }  Y^{\mu \alpha}_{e b  } X_{c d } - \frac{1}{24}   B^{\alpha \alpha}_{a b | c d }Y^{\mu\nu}_{ ae  }  Y^{\mu\nu}_{e b }  X_{c d } \nn
 & + \frac{1}{2} B^{\mu \nu}_{a b | c d } (D_\mu X)_{a c } (D_\nu X)_{b d } +
\frac{1}{18}B^{\mu \nu}_{a b | c d }  (D_\alpha Y^{\alpha \mu})_{a c } (D_\beta Y^{\beta \nu})_{b d }  + \frac{1}{6}B^{\mu \nu}_{a b | c d } (D_\mu X)_{ac} (D_\beta Y^{\beta \nu})_{b d } \Biggr] \ .
\label{eq:2loopBct}
\end{align}
The graphs factor into two one-loop integrals, so the counterterms can be determined by using eq~\eqref{3.11a} for each loop, and including combinatorial factors for grouping four $\eta$ fields into two groups of two fields each.

\subsection{Cubic $A$ terms}\label{sec:cubic}

The two-loop graphs from the sunset topology in Fig.~\ref{fig:two}(a) plus the associated subdivergence graphs give local counterterms. The symmetry relation Eq.~\eqref{3.9} is used to simplify the final results, and put the counterterm operators in a standard form. The possible counterterms are shown in Table~\ref{tab:ctAstructures}. To avoid integration by parts ambiguities, we choose to eliminate all derivatives on $X$ and $Y_{\mu \nu}$, so the derivatives act only on the $A$ vertices. Many flavor contractions can be eliminated by systematically applying Eq.~\eqref{3.9}. We give one example, and leave the analysis of the other cases to the reader. The identity Eq.~\eqref{3.9} is represented graphically in Fig.~\ref{fig:iden}.
%
%
\begin{figure}
\begin{center}
\begin{tikzpicture}
\filldraw (0,0) circle (0.05);
\draw (0,0) -- (60:1);
\draw (0,0) -- (0:1);
\draw (0,0) -- (-60:1);
\draw (60:0.35) circle (0.05) ;
\draw (2,0) node {$+$};
\begin{scope}[shift = {(3,0)}]
\filldraw (0,0) circle (0.05);
\draw (0,0) -- (60:1);
\draw (0,0) -- (0:1);
\draw (0,0) -- (-60:1);
\draw (0:0.35) circle (0.05) ;
\draw (2,0) node {$+$};
\end{scope}
\begin{scope}[shift = {(6,0)}]
\filldraw (0,0) circle (0.05);
\draw (0,0) -- (60:1);
\draw (0,0) -- (0:1);
\draw (0,0) -- (-60:1);
\draw (-60:0.35) circle (0.05) ;
\draw (2,0) node {$=0$};
\end{scope}
\end{tikzpicture}
\end{center}
\caption{\label{fig:iden} Graphical representation of the symmetry relation $A^\mu_{a | b c} + A^\mu_{b | c a} + A^\mu_{c | a b} = 0$. The lines from top to bottom correspond to the flavor indices $a,b,c$, and the line with the open circle denotes the first index of $A^\mu$, which is contracted with $(D_\mu \eta)$.}
\end{figure}
%
%
Consider a two-loop graph with two $A^\mu$ vertices, where $X$ is inserted on one $\eta$ line, and $Y_{\mu \nu}$ on a different $\eta$ line.
%
%
\begin{figure}
\begin{center}
\begin{align*}
\begin{array}{ccc}
\begin{tikzpicture}[
mid/.style = {decoration={
    markings,mark=at position 0.53 with {\arrow{latex}}}}]
\draw circle (0.75);
\filldraw (0:0.75) circle (0.05);
\filldraw (180:0.75) circle (0.05);
\filldraw (90:0.75) circle (0.05);
\filldraw (-90:0.75) circle (0.05);
\draw (150:0.75) circle (0.05);
\draw (30:0.75) circle (0.05);
\draw (-0.75,0) -- (0.75,0) ;
\draw (90:1.15) node {$X$};
\draw (-90:1.15) node {$Y$};
%
\end{tikzpicture}
&
\begin{tikzpicture}[
mid/.style = {decoration={
    markings,mark=at position 0.53 with {\arrow{latex}}}}]
\draw circle (0.75);
\filldraw (0:0.75) circle (0.05);
\filldraw (180:0.75) circle (0.05);
\filldraw (90:0.75) circle (0.05);
\filldraw (-90:0.75) circle (0.05);
\draw (150:0.75) circle (0.05);
\draw (0.5,0) circle (0.05);
\draw (-0.75,0) -- (0.75,0) ;
\draw (90:1.15) node {$X$};
\draw (-90:1.15) node {$Y$};
\end{tikzpicture}
&
%
\begin{tikzpicture}[
mid/.style = {decoration={
    markings,mark=at position 0.53 with {\arrow{latex}}}}]
\draw circle (0.75);
\filldraw (0:0.75) circle (0.05);
\filldraw (180:0.75) circle (0.05);
\filldraw (90:0.75) circle (0.05);
\filldraw (-90:0.75) circle (0.05);
\draw (150:0.75) circle (0.05);
\draw (-30:0.75) circle (0.05);
\draw (-0.75,0) -- (0.75,0) ;
\draw (90:1.15) node {$X$};
\draw (-90:1.15) node {$Y$};
\end{tikzpicture}
\\
\begin{tikzpicture}[
mid/.style = {decoration={
    markings,mark=at position 0.53 with {\arrow{latex}}}}]
\draw circle (0.75);
\filldraw (0:0.75) circle (0.05);
\filldraw (180:0.75) circle (0.05);
\filldraw (90:0.75) circle (0.05);
\filldraw (-90:0.75) circle (0.05);
\draw (30:0.75) circle (0.05);
\draw (-0.5,0) circle (0.05);
\draw (-0.75,0) -- (0.75,0) ;
\draw (90:1.15) node {$X$};
\draw (-90:1.15) node {$Y$};
\end{tikzpicture}
&
%
\begin{tikzpicture}[
mid/.style = {decoration={
    markings,mark=at position 0.53 with {\arrow{latex}}}}]
\draw circle (0.75);
\filldraw (0:0.75) circle (0.05);
\filldraw (180:0.75) circle (0.05);
\filldraw (90:0.75) circle (0.05);
\filldraw (-90:0.75) circle (0.05);
\draw (0.5,0) circle (0.05);
\draw (-0.5,0) circle (0.05);
\draw (-0.75,0) -- (0.75,0) ;
\draw (90:1.15) node {$X$};
\draw (-90:1.15) node {$Y$};
%
\end{tikzpicture}
&
\begin{tikzpicture}[
mid/.style = {decoration={
    markings,mark=at position 0.53 with {\arrow{latex}}}}]
\draw circle (0.75);
\filldraw (0:0.75) circle (0.05);
\filldraw (180:0.75) circle (0.05);
\filldraw (90:0.75) circle (0.05);
\filldraw (-90:0.75) circle (0.05);
\draw (-30:0.75) circle (0.05);
\draw (-0.5,0) circle (0.05);
\draw (-0.75,0) -- (0.75,0) ;
\draw (90:1.15) node {$X$};
\draw (-90:1.15) node {$Y$};
\end{tikzpicture}
\\
\begin{tikzpicture}[
mid/.style = {decoration={
    markings,mark=at position 0.53 with {\arrow{latex}}}}]
\draw circle (0.75);
\filldraw (0:0.75) circle (0.05);
\filldraw (180:0.75) circle (0.05);
\filldraw (90:0.75) circle (0.05);
\filldraw (-90:0.75) circle (0.05);
\draw (210:0.75) circle (0.05);
\draw (30:0.75) circle (0.05);
\draw (-0.75,0) -- (0.75,0) ;
\draw (90:1.15) node {$X$};
\draw (-90:1.15) node {$Y$};
\end{tikzpicture}
&
%
\begin{tikzpicture}[
mid/.style = {decoration={
    markings,mark=at position 0.53 with {\arrow{latex}}}}]
\draw circle (0.75);
\filldraw (0:0.75) circle (0.05);
\filldraw (180:0.75) circle (0.05);
\filldraw (90:0.75) circle (0.05);
\filldraw (-90:0.75) circle (0.05);
\draw (210:0.75) circle (0.05);
\draw (0.5,0) circle (0.05);
\draw (-0.75,0) -- (0.75,0) ;
\draw (90:1.15) node {$X$};
\draw (-90:1.15) node {$Y$};
\end{tikzpicture}
&
\begin{tikzpicture}[
mid/.style = {decoration={
    markings,mark=at position 0.53 with {\arrow{latex}}}}]
\draw circle (0.75);
\filldraw (0:0.75) circle (0.05);
\filldraw (180:0.75) circle (0.05);
\filldraw (90:0.75) circle (0.05);
\filldraw (-90:0.75) circle (0.05);
\draw (210:0.75) circle (0.05);
\draw (-30:0.75) circle (0.05);
\draw (-0.75,0) -- (0.75,0) ;
\draw (90:1.15) node {$X$};
\draw (-90:1.15) node {$Y$};
%
\end{tikzpicture}
\end{array}
\end{align*}
\end{center}
\caption{\label{fig:array} Allowed flavor contractions for $A^\mu A^\nu X Y$ with $X$ and $Y$ on different lines.}
\end{figure}
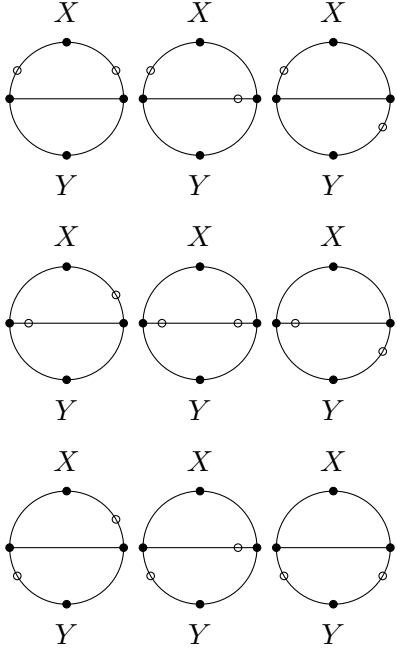
%
%
The allowed contractions are shown in Fig.~\ref{fig:array}. In each figure, the left $\eta^3$ vertex is $A^\mu$ and the right vertex is $A^\nu$, and the lines show the flavor index contractions, which generate operators of the form $A^\mu A^\nu X Y_{\mu \nu}$. The possible contractions  shown in the figure can be denoted as a matrix
\begin{align}
\begin{bmatrix} c_{11} & c_{12} & c_{13} \\  c_{21} & c_{22} & c_{23} \\  c_{31} & c_{32} & c_{33} \end{bmatrix} \,.
\label{4.9}
\end{align}
We can switch the two $\eta^3$ vertices in each figure. This swaps the Lorentz indices $\mu,\nu$, and swaps the flavor indices in $X$ and $Y_{\mu \nu}$. Both $X$ and $Y_{\mu \nu}$ are symmetric under the combined swap of Lorentz and flavor indices. Thus the matrix is symmetric, $c_{ab}=c_{ba}$, because transposing the graph transposes the location of the open circle. The symmetry relation Eq.~\eqref{3.9} implies that each row and each column of the matrix add to zero, as can be seen from Fig.~\ref{fig:iden}. The elements of a $3\times3$ symmetric matrix with each row and column adding to zero can all be determined from the three diagonal elements $c_{ii}$, so we write the 9 possible contractions in terms of the three independent ones. One can similarly analyze the other cases. Two structures in Table~\ref{tab:ctAstructures} are not allowed. $AAY_{\mu \nu}$ vanishes because the only Lorentz index contraction is $AAY_{\mu \mu}$. $A^\mu A D^3$ vanishes because the only flavor index contraction is $A_{abc} A^\mu_{a | b c}$ which vanishes using the symmetry relation Eq.~\eqref{3.9}.\footnote{Covariant derivatives maintain the symmetry properties of a tensor.}

The independent counterterms are
\begin{align}
\cL^{(A,2)}_{\text{c.t.}} &=\frac{1}{(16 \pi^2)^2 }\biggl[ a_{1,1} D_\mu A_{a b c } D_\mu A_{a b c } + a_{2,1} A_{a b c } X_{c d } A_{a b d } \nn
& + a_{3,1} D_\mu A^\mu_{a | b c}  A_{a b d } X_{c d } + a_{3,2} A^\mu_{a | b c}  D_\mu A_{a b d } X_{c d } + a_{4,1} D_\nu A^\mu_{a | b c}  A_{a b d } Y^{\mu \nu}_{c d } + a_{4,2} A^\mu_{a | b c }  D_\nu  A_{a b d } Y^{\mu \nu}_{c d } \nn
& + a_{5,1} D^2 A^\mu_{a | b c } D^2 A^\mu_{a | b c } + a_{5,2} D_\alpha D_\mu A^\mu_{a | b c } D_\alpha D_\nu A^\nu_{a | b c } \nn
& + a_{6,1} D^2 A^\mu_{a | b c } A^\mu_{a | b d } X_{c d } + a_{6,2} D^2 A^\mu_{c | a b } A^\mu_{d  | a b } X_{c d }  + a_{6,3} D_\alpha A^\mu_{a | b c } D_\alpha A^\mu_{a | b d } X_{c d }  + a_{6,4} D_\alpha A^\mu_{c | a b } D_\alpha A^\mu_{d  | a b } X_{c d } \nn
& + a_{6,5} D_\mu A^\mu_{a | b c } D_\nu A^\nu_{a | b d } X_{c d } + a_{6,6} D_\mu A^\mu_{c | a b } D_\nu A^\nu_{d  | a b } X_{c d }  + a_{6,7} D_\nu A^\mu_{a | b c } D_\mu A^\nu_{a | b d } X_{c d } \nn
& + a_{6,8} D_\nu A^\mu_{c | a b } D_\mu A^\nu_{d  | a b } X_{c d }  + a_{6,9} D_\nu D_\mu A^\mu_{a | b c }  A^\nu_{a | b d } X_{c d } + a_{6,10} D_\nu D_\mu A^\mu_{c | a b }  A^\nu_{d | a b } X_{c d } \nn
& + a_{7,1} D_\alpha A^\mu_{a | b c } D_\alpha A^\nu_{a | b d } Y^{\mu \nu}_{cd} + a_{7,2} D_\alpha A^\mu_{c | a b } D_\alpha A^\nu_{d  | ab} Y^{\mu \nu}_{cd}  + a_{7,3} D_\mu A^\alpha_{a | b c} D_\nu A^\alpha_{a | b d} Y^{\mu \nu}_{cd} \nn
& + a_{7,4} D_\mu A^\alpha_{c | a b } D_\nu A^\alpha_{d  | ab} Y^{\mu \nu}_{cd} + a_{7,5} D_\mu A^\alpha_{a | b c} D_\nu A^\nu_{a | b d} Y^{\mu \alpha}_{cd} + a_{7,6} D_\mu A^\alpha_{c | a b } D_\nu A^\nu_{d  | ab} Y^{\mu \alpha}_{cd}  \nn
& + a_{7,7} D_\nu A^\alpha_{a | b c} D_\mu A^\nu_{a | b d} Y^{\mu \alpha}_{cd} + a_{7,8} D_\nu A^\alpha_{c | a b } D_\mu A^\nu_{d  | ab} Y^{\mu \alpha}_{cd}  + a_{7,9} A^\alpha_{a | b c} D_\mu D_\nu  A^\nu_{a | b d} Y^{\mu \alpha}_{cd} \nn
& + a_{7,10}  A^\alpha_{c | a b } D_\mu D_\nu  A^\nu_{d  | ab} Y^{\mu \alpha}_{cd} + a_{7,11} D_\mu  D_\nu A^\alpha_{a | b c} A^\nu_{a | b d} Y^{\mu \alpha}_{cd} + a_{7,12} D_\mu  D_\nu A^\alpha_{c | a b } A^\nu_{d  | ab} Y^{\mu \alpha}_{cd} \nn
& + a_{8,1} A^\mu_{c | ab } A^\mu_{d  | a b } X_{ce} X_{ed} + a_{8,2} A^\mu_{a | b c } A^\mu_{ a | b d } X_{ce} X_{ed} + a_{8,3} A^\mu_{a | b c } A^\mu_{ e | b d } X_{ae} X_{cd} + 
a_{8,4} A^\mu_{a | b c } A^\mu_{ a | d e } X_{bd} X_{ce} \nn
& + a_{9,1} A^\mu_{c | ab } A^\nu_{d  | a b } ( X_{ce} Y^{\mu \nu}_{ed} + Y^{\mu \nu}_{ce} X_{ed} ) + a_{9,2} A^\mu_{a | b c } A^\nu_{ a | b d } ( X_{ce} Y^{\mu \nu}_{ed} + Y^{\mu \nu}_{ce} X_{ed}) \nn
&+ a_{9,3} A^\mu_{a | b c } A^\nu_{ e | b d } X_{ae} Y^{\mu \nu}_{cd}  + a_{9,4} A^\mu_{a | b c } A^\nu_{ a | d e } X_{ce} Y^{\mu \nu}_{bd} + a_{9,5} A^\mu_{a | b c } A^\nu_{ e | b d } X_{cd} Y^{\mu \nu}_{ae} \nn
& +  a_{10,1} A^\mu_{c | ab } A^\mu_{d  | a b } Y^{\alpha \beta}_{ce} Y^{\alpha \beta}_{ed} + 
a_{10,2} A^\mu_{a | b c } A^\mu_{ a | b d } Y^{\alpha \beta}_{ce} Y^{\alpha \beta}_{ed} 
+ a_{10,3} A^\mu_{c | ab } A^\nu_{d  | a b } Y^{\mu \alpha}_{ce} Y^{\nu \alpha}_{ed}  \nn
& + 
a_{10,4} A^\mu_{a | b c } A^\nu_{ a | b d } Y^{\mu \alpha}_{ce} Y^{\nu \alpha}_{ed} + 
a_{10,5} A^\mu_{c | ab } A^\nu_{d  | a b } Y^{\nu \alpha}_{ce} Y^{\mu \alpha}_{ed} + 
a_{10,6} A^\mu_{a | b c } A^\nu_{ a | b d } Y^{\nu \alpha}_{ce} Y^{\mu \alpha}_{ed} \nn
& + 
a_{10,7} A^\mu_{a | b c } A^\mu_{ e | b d } Y^{\alpha \beta}_{ae} Y^{\alpha \beta}_{cd} +
a_{10,8} A^\mu_{a | b c } A^\mu_{ a | d e } Y^{\alpha \beta}_{bd} Y^{\alpha \beta}_{ce} +
a_{10,9} A^\mu_{a | b c } A^\nu_{ e | b d } ( Y^{\mu \alpha}_{ae} Y^{\nu \alpha}_{cd} + Y^{\nu \alpha}_{ae} Y^{\mu \alpha}_{cd}) \nn
& + 
a_{10,10}A^\mu_{a | b c } A^\nu_{ a | d e } (Y^{\mu \alpha}_{bd} Y^{\nu \alpha}_{ce} + Y^{\nu \alpha}_{bd} Y^{\mu \alpha}_{ce}) +
a_{10,11} A^\mu_{a | b c } A^\nu_{ b | e d } (Y^{\mu \alpha}_{ae} Y^{\nu \alpha}_{cd} - Y^{\nu \alpha}_{ae} Y^{\mu \alpha}_{cd}) \biggr]\,.
\label{eq:2loopAct}
\end{align}
There are two additional $A^\mu A^\nu Y D^2$ operators
\begin{align}
& a_{7,13} D^2 A^\mu_{a | b c} A^\nu_{a | b d} Y^{\mu \nu}_{cd} + a_{7,14} D^2 A^\mu_{c | a b } A^\nu_{d  | ab} Y^{\mu \nu}_{cd} .
\label{28.16}
\end{align}
These can be written in terms of the other $a_{7}$ operators and $a_{10,11}$ using the Bianchi identity
\begin{align}
D_\alpha Y_{\beta \gamma} + D_\beta  Y_{\gamma \alpha} + D_\gamma Y_{\alpha \beta} = 0 \,.
\end{align}
The coefficients of the two-loop counterterms are determined from the sunset graphs in Fig.~\ref{fig:two}(a), and are listed in Table~\ref{tab:aact}.

The calculation of Eq.~\eqref{eq:2loopAct} involves the evaluation of at most four-point Feynman integrals. These are sufficient to compute the two-loop running of EFT operators with an arbitrary number of external legs. The usual diagrammatic computation requires computing higher point graphs, which is more difficult.

\begin{table}
\renewcommand{\arraycolsep}{0.25cm}
\renewcommand{\arraystretch}{1.25}
\begin{align*}
\begin{array}{l | l | l | l}
	a_{1,1} = -\frac{3}{4 \epsilon} , &
	a_{2,1} =  \frac{9}{2\epsilon^2}  - \frac{9}{2\epsilon}, & & \\[5pt] \hline
	a_{3,1} = \frac{3}{2 \epsilon ^2}  -\frac{15}{4 \epsilon }, &
	a_{3,2} =  \frac{9}{2 \epsilon ^2}  -\frac{9}{4 \epsilon }, &
	a_{4,1} =  -\frac{3}{2 \epsilon ^2}  +\frac{7}{4 \epsilon }, &
	a_{4,2} =  -\frac{3}{2 \epsilon ^2}  -\frac{5}{4 \epsilon }, \\[5pt] \hline
	a_{5,1} =  \frac{1}{64 \epsilon }, & 
	a_{5,2} =  -\frac{1}{48 \epsilon }, & &  \\ \hline
	a_{6,1} =  \frac{1}{36 \epsilon ^2}+\frac{25}{216 \epsilon }, & 
	a_{6,2} =  \frac{13}{72 \epsilon ^2}-\frac{107}{432 \epsilon }, &
	a_{6,3} =  -\frac{5}{36 \epsilon ^2}+ \frac{37}{216 \epsilon }, &
	a_{6,4} =  \frac{2}{9 \epsilon ^2}-\frac{2}{27 \epsilon }, \\[5pt]
	a_{6,5}=  \frac{1}{36 \epsilon ^2}-\frac{5}{216 \epsilon }, &
	a_{6,6}=  -\frac{5}{72 \epsilon ^2}-\frac{65}{432 \epsilon }, &
	a_{6,7}=  \frac{1}{36 \epsilon ^2}-\frac{5}{216 \epsilon }, &
	a_{6,8}=  \frac{13}{72 \epsilon ^2}-\frac{11}{432 \epsilon }, \\[5pt]
	a_{6,9}=  -\frac{1}{9 \epsilon ^2}+ \frac{5}{54 \epsilon }, & 
	a_{6,10}= \frac{1}{36 \epsilon ^2}-\frac{59}{216 \epsilon }, & &  \\[5pt] \hline
	a_{7,1}=  -\frac{1}{48 \epsilon }, &
	a_{7,2}=  -\frac{13}{96 \epsilon }, &
	a_{7,3}=  \frac{1}{18 \epsilon ^2}+\frac{1}{432 \epsilon}, &
	a_{7,4}=  -\frac{1}{72 \epsilon ^2}-\frac{41}{864 \epsilon }, \\[5pt]
	a_{7,5}=  -\frac{1}{36 \epsilon ^2}+\frac{13}{432 \epsilon }, &
	a_{7,6}=  \frac{5}{72 \epsilon ^2}-\frac{191}{864 \epsilon }, &
	a_{7,7}=  \frac{1}{36 \epsilon ^2}-\frac{13}{432 \epsilon }, &
	a_{7,8}=  \frac{13}{72 \epsilon ^2}-\frac{61}{864 \epsilon }, \\[5pt]
	a_{7,9}=  -\frac{1}{36 \epsilon ^2}-\frac{17}{432 \epsilon },&
	a_{7,10}= \frac{5}{72 \epsilon ^2}-\frac{149}{864 \epsilon }, & 
	a_{7,11}=   \frac{1}{36 \epsilon ^2}-\frac{19}{432 \epsilon }, &
	a_{7,12}=  \frac{13}{72 \epsilon ^2}-\frac{139}{864 \epsilon }, \\[5pt] \hline
	a_{8,1}=  -\frac{5}{16 \epsilon ^2}+\frac{19}{96\epsilon }, &
	a_{8,2}=  \frac{1}{8 \epsilon ^2} -\frac{11}{48 \epsilon }, &
	a_{8,3}=  -\frac{1}{4 \epsilon ^2}+\frac{5}{8 \epsilon}, & 
	a_{8,4}= -\frac{1}{2 \epsilon ^2}+ \frac{1}{8 \epsilon }, \\[5pt] \hline
	a_{9,1}= \frac{13}{72 \epsilon ^2}-\frac{11}{432 \epsilon },&
	a_{9,2}=  \frac{1}{36 \epsilon ^2}-\frac{5}{216 \epsilon }, &
	a_{9,3}=  -\frac{19}{36 \epsilon ^2}+ \frac{5}{216 \epsilon}, & 
	a_{9,4}=  \frac{11}{36 \epsilon ^2}+\frac{17}{216 \epsilon }, \\[5pt]
	a_{9,5}=  \frac{11}{36 \epsilon ^2}-\frac{145}{216 \epsilon }, & & & \\[5pt] \hline
	a_{10,1}= \frac{35}{1152 \epsilon }-\frac{5}{96 \epsilon ^2}, &
	a_{10,2}=  \frac{1}{48 \epsilon ^2}-\frac{25}{576 \epsilon }, &
	a_{10,3}= \frac{13}{144 \epsilon^2}+\frac{251}{1728 \epsilon } &
	a_{10,4}=  \frac{1}{72 \epsilon ^2}+\frac{11}{864 \epsilon }, \\[5pt]
	a_{10,5}=  \frac{13}{144 \epsilon ^2}-\frac{217}{1728 \epsilon }, &
	a_{10,6}=  \frac{1}{72 \epsilon ^2}-\frac{25}{864 \epsilon }, &
	a_{10,7}=  \frac{1}{72 \epsilon ^2}-\frac{67}{864 \epsilon }, & 
	a_{10,8}=  \frac{1}{36 \epsilon ^2}-\frac{25}{1728 \epsilon },  \\[5pt]
	a_{10,9}=  -\frac{29}{144 \epsilon}, &
	a_{10,10}=  \frac{19}{288 \epsilon }, &
	a_{10,11}=  -\frac{1}{8 \epsilon }
\end{array}
\end{align*}
\caption{\label{tab:aact} Table of coefficients for the two-loop counterterms.}
\end{table}

\newpage
\section{Factorizable Topologies}\label{sec:factorizable}

A factorizable graph $G$ is a graph which is the union of two subgraphs, $G=G_1 \cup G_2$, where the intersection of the two subgraphs $G_1 \cap G_2$ is a single vertex, and each subgraph contains at least one loop.
The two subgraphs in a factorizable graph do not have any propagators in common. The Feynman integral for such graphs can be written as the product of the Feynman integrals for each subgraph.
The two-loop $B$-graphs are factorizable graphs, give purely $1/\epsilon^2$ counterterms in the MS scheme, and do not contribute to the two-loop anomalous dimension. This is a general feature of factorizable diagrams in the MS scheme;  they can be  omitted if one is only interested in computing the RGE of the theory. Consider the general graph Fig.~\ref{fig:two}(b) with arbitrary insertions of the external field vertices, which can insert momentum. Let $k$ and $l$ be the loop momenta, $p_1,\ldots , p_r$ the incoming momenta from vertices on the $k$ loop, and $q_1,\ldots,q_s$ the incoming momenta from vertices on the $l$ loop. The two-loop integral has the product form
\begin{align}
I &= I^{\{\alpha\}}_1(k,\{p\}) \ I^{\{\alpha\}}_2(l,\{q\}) \,,
\label{4.2}
\end{align}
where momentum conservation implies $\sum p_r + \sum q_s = 0$. The superscript $\{\alpha\}$ denotes possible  indices contractions between the two loops. $I_1$ only depends on the momenta in $I_2$ through the momentum conservation relation, and does not depend on the momenta of the individual lines in $I_2$. Each graph integral can be evaluated separately, and has a divergent piece, denoted as $I_\infty/\epsilon$, and a (possibly non-local) finite piece, denoted as $I_f$. The full integral $I$ reads
\begin{align}
I &=\left[ \frac{1}{\epsilon} I^{\{\alpha\}}_{1\infty} (\{p\}) +  I^{\{\alpha\}}_{1f}(\{p\}) \right] \left[ \frac{1}{\epsilon} I^{\{\alpha\}}_{2 \infty} (\{q\}) +  I^{\{\alpha\}}_{2f}(\{q\}) \right] \,.
\label{4.3}
\end{align}
The two-loop graph integral has two possible one-loop subdivergences. Because of the product form Eq.~\eqref{4.2}, the subdivergence in each loop is exactly equal to $I_{\infty}$. The subdivergence subtraction is
\begin{align}
I_{\text{sub}} &= - \left[ \frac{1}{\epsilon} I^{\{\alpha\}}_{1\infty} (\{p\})  \right] \left[ \frac{1}{\epsilon} I^{\{\alpha\}}_{2 \infty} (\{q\}) +  I^{\{\alpha\}}_{2f}(\{q\}) \right] 
- \left[ \frac{1}{\epsilon} I^{\{\alpha\}}_{1\infty} (\{p\}) +  I^{\{\alpha\}}_{1f}(\{p\}) \right] \left[ \frac{1}{\epsilon} I^{\{\alpha\}}_{2 \infty} (\{q\}) \right] \,,
\label{4.4}
\end{align}
and the subdivergence subtracted two-loop graph integral is
\begin{align}
I_{\rm tot} \equiv I+I_{\rm sub}&= - \frac{1}{\epsilon^2} I^{\{\alpha\}}_{1\infty} (\{p\}) I^{\{\alpha\}}_{2 \infty} (\{q\}) +  I^{\{\alpha\}}_{1f}(\{p\}) I^{\{\alpha\}}_{2f}(\{q\})\,.
\label{4.5}
\end{align}
Notice that the $1/\epsilon$ pieces exactly cancel. Therefore, the two-loop counterterm is purely $1/\epsilon^2$,
\begin{align}
I_{\text{c.t.}} &=  \frac{1}{\epsilon^2} I^{\{\alpha\}}_{1\infty} (\{p\})\ I^{\{\alpha\}}_{2 \infty} (\{q\}) \,,
\label{4.6}
\end{align}
and equals the product of the individual counterterms $-I_\infty/\epsilon$ of each subgraph. The finite part is given by the product of the finite parts of the individual subgraphs,
\begin{align}
I_{f} &= I^{\{\alpha\}}_{1f} (\{p\})\ I^{\{\alpha\}}_{2 f} (\{q\}) \,.
\label{4.7}
\end{align}
The argument is generalizable to factorizable graphs with arbitrary loops. Consider first a fully factorizable $L$ graph, i.e.\ one which factors into $L$ one-loop subgraphs. The divergence of the subtracted $L$ loop graph is $(-1)^{L-1}$ times the individual divergences, and is purely $1/\epsilon^L$. The finite part is the product of the finite parts of the individual subgraphs, and there are no $1/\epsilon^k$ pieces for $k \ge 1, k \not =L$,
\begin{align}
I_{\rm tot} &= (-1)^{L-1} \frac{1}{\epsilon^L} I^{\{\alpha\}}_{1\infty} \ldots I^{\{\alpha\}}_{L\infty}   +  I^{\{\alpha\}}_{1f} \ldots I^{\{\alpha\}}_{Lf} \,, \nn
I_{\text{c.t.}} &= (-1)^{L} \frac{1}{\epsilon^L} I^{\{\alpha\}}_{1\infty} \ldots I^{\{\alpha\}}_{L\infty}  \,.
\label{4.8}
\end{align}
The argument also applies to partially factorizable graphs, which factor into $n_{\rm nf}$  non-factorizable subgraphs, which can each have more than one loop. The proof is the same as for fully factorizable graphs where one or more of the simple one-loop parts are replaced by subdivergence subtracted (non-factorizable) higher-loop graphs. In this case, for a subtracted $L$ loop graph, the lowest pole is not $1/\epsilon^L$, but instead $1/\epsilon^{n_{\rm nf}}$ where $n_{\rm nf}$ is the number of non-factorizable parts ($n_{\rm nf}=1$ for a non-factorizable graph). 
Thus factorizable topologies do not contribute to the anomalous dimensions, which depend only on the $1/\epsilon$ pole.
%
%
%
\begin{figure}
\begin{center}
\begin{tikzpicture}[scale=0.6]
\draw (-1,0) circle (1);
\draw (1,0) circle (1);
\filldraw (0,0) circle (0.05);
\filldraw (-2,0) circle (0.05);
\filldraw (2,0) circle (0.05);
\draw (2,0) -- +(45:1.5);
\draw (2,0) -- +(-45:1.5);
\draw (-2,0) -- +(135:1.5);
\draw (-2,0) -- +(225:1.5);
\draw (0,-2.5) node {(a)};
\end{tikzpicture}
\hspace{2cm}
\begin{tikzpicture}[scale=0.6]
\draw (1,0) circle (0.5 and 1);
\draw (-2,0) -- (1,1);
\draw (-2,0) -- (1,-1);
\draw (-2,0) -- +(135:1.5);
\draw (-2,0) -- +(225:1.5);
\draw (1,1) -- (2,2);
\draw (1,-1) -- (2,-2);
\filldraw (1,1) circle (0.05);
\filldraw (1,-1) circle (0.05);
\filldraw (-2,0) circle (0.05);
\draw (0,-2.5) node {(b)};
\end{tikzpicture}
\end{center}
\caption{Two-loop graphs renormalizing the $\phi^4$ interaction. Graph (a) does not contribute to the anomalous dimension of $\lambda$.
\label{fig:lambda}}
\end{figure}
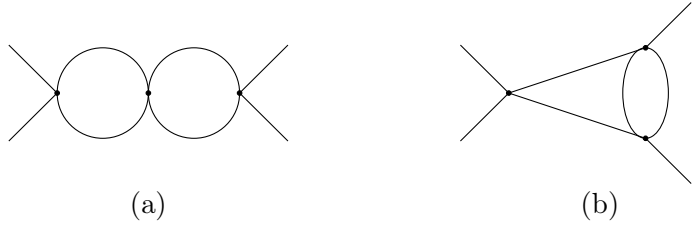
%
%

For example, the anomalous dimension for the scalar quartic coupling in a general renormalizable theory was computed in Ref.~\cite{Machacek:1984zw}. There are two topologies of scalar loops for the quartic term, shown in Fig.~\ref{fig:lambda}. Graph (a) has a factorizable topology, and should not contribute to the anomalous dimension according to the argument above, while graph (b) should contribute. To see this explicitly, let us compare the flavor contractions appearing in each graph. For a $\phi^4$ coupling defined as
\begin{align}
\mathcal{L} &= - \frac{1}{4!} \lambda_{abcd}\, \phi^a \phi^b \phi^c \phi^d \,,
\end{align}
the type of flavor contractions from graph (a) is $\lambda_{abef} \lambda_{efgh} \lambda_{ghcd}$, and from graph (b) is $\lambda_{abef} \lambda_{cegh} \lambda_{dfgh}$. The first type is absent from the RGE for $\lambda_{abcd}$ given in Ref.~\cite[(4.3)]{Machacek:1984zw}, showing indeed that graph (a) does not contribute to the RGE of the quartic coupling.
Usually, Feynman graphs are grouped by powers of coupling constant, and both graphs in Fig.~\ref{fig:lambda} are $\mathcal{O}(\lambda^3)$. The method of this paper groups the graphs by topology rather than coupling constant, which makes the absence of some type of flavor contractions manifest.

There is one subtlety in the above argument --- evanescent operators~\cite{Buras:1989xd,Dugan:1990df,Herrlich:1994kh}. Consider the two-loop case where $I_{1,2}$ each have two Lorentz indices. If $I_1^{\alpha_1\alpha_2}$ and $I_2^{\alpha_1\alpha_2}$ have terms proportional to $\eta_{\alpha_1 \alpha_2}$, then their product gives terms of the form $\eta_{\alpha \alpha}=d=4-2\epsilon$. The order $\epsilon$ piece in the Lorentz contraction multiplied by the $1/\epsilon^2$ divergence from the integral leads to a $1/\epsilon$ piece. 
In our calculation, the only terms where this happens are $B^{\alpha \alpha}$ terms when $B^{\mu\nu}$ is proportional to $\eta^{\mu\nu}$. Here the $\eta_{\alpha \alpha}$ contraction arises from one of the loops in Fig.~\ref{fig:two}(b), when both internal $\eta$ lines are the $D\eta$ fields in Eq.~\eqref{3.7}, $\vev{D_\mu \eta D_\nu \eta} \to \eta_{\mu \nu}$. In this case, the one-loop subgraph has the form
\begin{align}
\eta_{\alpha \alpha} \left[ \frac{1}{\epsilon} I_{1\infty} (\{p\}) +   I_{1f}(\{p\}) \right] =  \left[ \frac{4}{\epsilon} I_{1\infty} (\{p\}) +  d\, I_{1f}(\{p\})
- 2  I_{1\infty} (\{p\})  \right] \,.
\label{7.11}
\end{align}
The usual minimal subtraction procedure subtracts the $4I_{\infty} (\{p\})/\epsilon$ term. The two-loop graph has the form
\begin{align}
I &=\eta_{\alpha \alpha} \left[ \frac{1}{\epsilon} I_{1\infty} (\{p\}) +  I_{1f}(\{p\}) \right] \left[ \frac{1}{\epsilon} I_{2 \infty} (\{q\}) +  I_{2f}(\{q\}) \right] \,.
\end{align}
Subtracting the subdivergences gives
\begin{align}
I_{\rm tot} &= \eta_{\alpha \alpha} \left[ \frac{1}{\epsilon} I_{1\infty} (\{p\}) +  I_{1f}(\{p\}) \right] \left[ \frac{1}{\epsilon} I_{2 \infty} (\{q\}) +  I_{2f}(\{q\}) \right] \nn
& -  \left[ \frac{4}{\epsilon} I_{1\infty} (\{p\}) \right] \left[ \frac{1}{\epsilon} I_{2 \infty} (\{q\}) +  I_{2f}(\{q\}) \right] \nn
&- d \left[ \frac{1}{\epsilon} I_{1\infty} (\{p\}) +  I_{1f}(\{p\}) \right] \left[ \frac{1}{\epsilon} I_{2 \infty} (\{q\})  \right] \nn
&= - \frac{4}{\epsilon^2} I_{1\infty} (\{p\})I_{2\infty} (\{p\}) + 4 I_{1f}(\{p\})I_{2f}(\{p\}) 
- 2 I_{1\infty} (\{p\})  I_{2f}(\{p\}) + \mathcal{O}(\epsilon) \,,
\end{align}
which has no $1/\epsilon$ term. Thus the $B$-terms in Eq.~\eqref{eq:2loopBct} do not contribute to the anomalous dimensions, and $\eta_{\alpha \alpha}$ pieces in $B_{\alpha \alpha}$ are taken to be $4$.

A more interesting case is when factors of $\epsilon$ are generated only after combining the loops. For example, if each loop produces a $\eta_{\mu \nu}$, then the factor of $d$ is produced only when the loops are combined, not in an individual loop. In this case, the factorizable topologies can produce $1/\epsilon$ terms. However, it is possible to remove them by an additional finite subtraction, analogous to that used for evanescent operators~\cite{Dugan:1990df}. This was noted in some examples in Refs.~\cite{Bern:2013qca,Bern:2020ikv}, and the argument is completely general.
Consider what happens when we apply the two-loop counterterm formula to a Lagrangian where we split the interaction part\footnote{The kinetic term must be kept in $d$ dimensions to regulate the loop integrals. All other terms can be included in $\mathcal{L}_\text{int} $. $B^{\mu\nu}$ only receives contributions from nonrenormalizable interactions and is part of $\mathcal{L}_\text{int} $.} into a four-dimensional and an evanescent part
\begin{equation}
\mathcal{L}_\text{int} = {\bar{\mathcal{L}}}_{\text{int}} +  {\widehat{\mathcal{L}}}_{\text{int}}  \,,
\end{equation}
where in ${\bar{\mathcal{L}}}_{\text{int}} $ all Lorentz indices take values in $0,..,3$ and $ {\widehat{\mathcal{L}}}_{\text{int}} $ contains evanescent operators. Insertions of ${\bar{\mathcal{L}}}_{\text{int}}$ into factorizable graphs will only generate $\bar{\eta}^{\mu\mu} = 4$. Equivalently, in Eq.~\eqref{eq:2loopBct}, ${\bar{\mathcal{L}}}_{\text{int}}$ will contribute to $B^{\mu\nu}$ only terms which are proportional to $\bar\eta^{\mu\nu}$. Therefore ${\bar{\mathcal{L}}}_{\text{int}}$ does not give rise to $1/\epsilon$ poles via factorizable graphs. Such factors of $\epsilon$ are now generated by insertions of evanescent operators through $\widehat \eta ^{\mu \mu} = -2 \epsilon$. But in a renormalization scheme where counterterms of physical operators are adjusted to compensate finite effects of evanescent insertions in addition to divergent parts, the evanescent operators do not contribute to the anomalous dimensions~\cite{Dugan:1990df}. The key observation in Ref.~\cite{Dugan:1990df} which makes this possible is that these additional pieces, analogous to the $- 2  I_{1\infty} (\{p\}) $ in Eq.~\eqref{7.11}, are \emph{local}, since they are proportional to divergent parts. As a result, in such a scheme factorizable graphs \textit{never} contribute to the anomalous dimensions. They also do not contribute in the evanescence-free scheme where the couplings of the evanescent operators are set to zero~\cite{Fuentes-Martin:2022vvu}.
Finally we note that even though the argument is most evident when one uses $\bar{\cL}_\text{int}$, it also holds in a scheme where operators are defined in $d$ dimensions, since that would only shift the coefficients of the evanescent operators.

\begin{figure}
\begin{center}
\begin{tikzpicture}[scale=0.75]
\draw(-1,0) circle (1);
\draw(1,0) circle (1);
\draw[decorate,decoration={snake}] (2,0) -- (3.5,0);
\draw[decorate,decoration={snake}] (-2,0) -- (-3.5,0);
\filldraw (2,0) circle (0.05);
\filldraw (-2,0) circle (0.05);
\end{tikzpicture}
\end{center}
\caption{\label{fig:penguin}. The double penguin graph due to the insertion of $(\overline \psi \gamma^\mu \psi)(\overline \psi \gamma_\mu \psi)$.}
\end{figure}
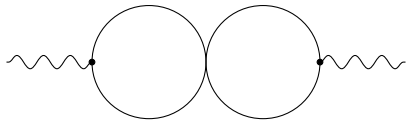

An explicit example calculation is the double-penguin graph in Ref.~\cite{Bern:2020ikv}, shown in Fig.~\ref{fig:penguin}, where the $\epsilon$ pieces arise from the trace of Dirac matrices in the two-loop diagram. Ref.~~\cite{Bern:2020ikv} showed that this graph does not contribute to the anomalous dimension using an additional finite subtraction beyond \msbar.
%
%
%
%

\begin{figure}
\begin{center}
\begin{tikzpicture}[scale=0.8]
\filldraw (-0.75,0) circle (0.05);
\filldraw (0.75,0) circle (0.05);
\draw (0.75,0) arc(0:360:0.75);
\draw (0.75,0) arc(0:180:0.75 and 0.3);
\draw (0.75,0) arc(0:-180:0.75 and 0.3);
\draw (0,-1.25) node {(a)};
\end{tikzpicture}
\hspace{1cm}
%
\begin{tikzpicture}[scale=0.8]
\draw (0,0) circle (0.75);
\filldraw (-90:0.75) circle (0.05);
\filldraw (30:0.75) circle (0.05);
\filldraw (150:0.75) circle (0.05);
\draw (-90:0.75) -- (30:0.75);
\draw (-90:0.75) -- (150:0.75);
\draw (0,-1.25) node {(b)};
\end{tikzpicture}
\hspace{1cm}
%
%
\begin{tikzpicture}[scale=0.8]
\draw (0,0) circle (0.75);
\draw (60:0.75) -- (-60:0.75);
\draw (120:0.75) -- (-120:0.75);
\filldraw (60:0.75) circle (0.05);
\filldraw (-60:0.75) circle (0.05);
\filldraw (120:0.75) circle (0.05);
\filldraw (-120:0.75) circle (0.05);
\draw (0,-1.25) node {(c)};
\end{tikzpicture}
\hspace{1cm}
%
%
\begin{tikzpicture}[scale=0.8]
\draw (0,0) circle (0.75);
\draw (0,0) -- (90:0.75);
\draw (0,0) -- (-30:0.75);
\draw (0,0) -- (210:0.75);
\filldraw (0,0) circle (0.05);
\filldraw (90:0.75) circle (0.05);
\filldraw (-30:0.75) circle (0.05);
\filldraw (210:0.75) circle (0.05);
\draw (0,-1.25) node {(d)};
\end{tikzpicture}

\vspace{0.5cm}

\begin{tikzpicture}[scale=0.8]
\filldraw (0,0) circle (0.05);
\draw [rotate=90] (0.5,0) circle (0.5 and 0.2);
\draw [rotate=-30] (0.5,0) circle (0.5 and 0.2);
\draw [rotate=210] (0.5,0) circle (0.5 and 0.2);
\draw (0,-1.25) node {(e)};
\end{tikzpicture}
\hspace{0.5cm}
%
\begin{tikzpicture}[scale=0.8]
\filldraw (-0.75,0) circle (0.05);
\filldraw (0.75,0) circle (0.05);
\draw (0,0) circle (0.75);
\draw (1.5,0) circle (0.75);
\draw (-1.5,0) circle (0.75);
\draw (0,-1.25) node {(f)};
\end{tikzpicture}
\hspace{0.5cm}
%
\begin{tikzpicture}[scale=0.8]
\draw (0.75,0) circle (0.75);
\draw (-0.75,0) circle (0.75);
\draw (0,0) -- (1.5,0);
\filldraw (1.5,0) circle (0.05);
\filldraw (0,0) circle (0.05);
\draw (0,-1.25) node {(g)};
\end{tikzpicture}
\hspace{0.5cm}
%
\begin{tikzpicture}[scale=0.8]
\draw (0.75,0) circle (0.75);
\draw (-0.75,0) circle (0.75);
\draw (0.75,0.75) -- (0.75,-0.75);
\filldraw (0.75,0.75) circle (0.05);
\filldraw (0.75,-0.75) circle (0.05);
\filldraw (0,0) circle (0.05);
\draw (0,-1.25) node {(h)};
\end{tikzpicture}
\end{center}
\caption{\label{fig:three} Skeleton graphs for three-loop corrections to the action. There can be an an arbitrary number of $\eta\eta$ vertex insertions, as in Fig.~\ref{fig:loop}.}
\end{figure}
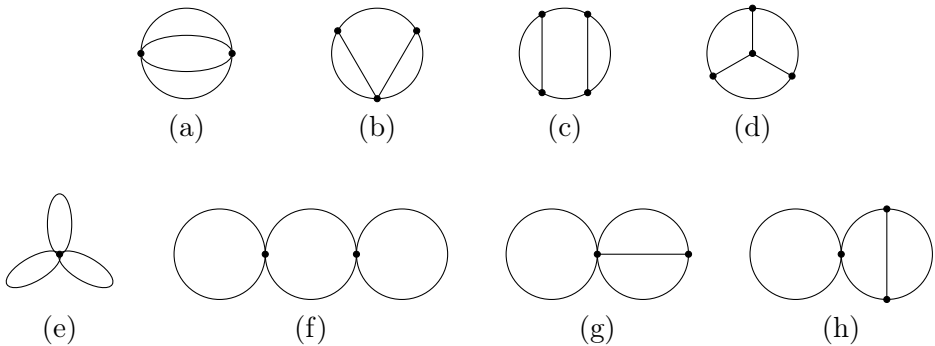

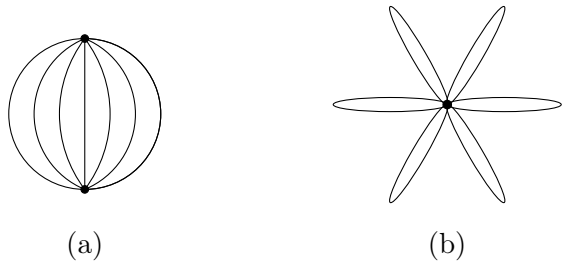
\begin{figure}
\begin{center}
\begin{tikzpicture}[scale=1]
\draw (0,0) circle (1);
\draw (0,-1) -- (0,1);
\filldraw (0,1) circle (0.05);
\filldraw (0,-1) circle (0.05);
\draw (1,0) arc(0:90:1);
\draw (1,0) arc(0:-90:1);
\draw (0.3333,0) arc(0:36.8699:1.66667);
\draw (0.3333,0) arc(0:-36.8699:1.66667);
\draw (0.6667,0) arc(0:67.3801:1.08333);
\draw (0.6667,0) arc(0:-67.3801:1.08333);
\draw[rotate=180] (0.3333,0) arc(0:36.8699:1.66667);
\draw[rotate=180] (0.3333,0) arc(0:-36.8699:1.66667);
\draw[rotate=180] (0.6667,0) arc(0:67.3801:1.08333);
\draw[rotate=180] (0.6667,0) arc(0:-67.3801:1.08333);
\draw (0,-1.75) node [align=center] {(a)};
\end{tikzpicture}
\hspace{2cm}
\begin{tikzpicture}[scale=0.75]
\draw[rotate=0] (0,0) arc(-180:180:1 and 0.125);
\draw[rotate=60] (0,0) arc(-180:180:1 and 0.125);
\draw[rotate=120] (0,0) arc(-180:180:1 and 0.125);
\draw[rotate=180] (0,0) arc(-180:180:1 and 0.125);
\draw[rotate=240] (0,0) arc(-180:180:1 and 0.125);
\draw[rotate=300] (0,0) arc(-180:180:1 and 0.125);
\filldraw (0,0) circle (0.075);
\draw (0,-2.5) node [align=center] {(b)};
\end{tikzpicture}
\end{center}
\caption{\label{fig:rge}. (a) $L$ loop non-factorizable graph with an $L+1$ field vertex. (b) $L$ loop factorizable graph with a $2L$-field vertex.}
\end{figure}

The graph factorization argument extends to arbitrary loop order. The skeleton three-loop graphs are shown in Fig.~\ref{fig:three}. The graphs (e,f,g,h) in Fig.~\ref{fig:three} do not contribute to the three-loop anomalous dimensions. The three-loop graphs in Fig.~\ref{fig:three} have vertices with up to 6 fields. If one is only interested in the RGEs, which depend on the non-factorizable graphs, then only vertices with up to 4 fields are needed. This makes the RGE computation much simpler than that of the full counterterm Lagrangian, which requires an expansion up to 6 fields. At four loops, the RGE requires vertices up to 5 fields, whereas the full counterterm calclulation requires an expansion up to 8 fields. In general at $L$ loops, one needs vertices  with up to $L+1$ fields for the RGE from graphs such as Fig.~\ref{fig:rge}(a), and vertices up to $2L$ fields for the full $L$ loop correction from graphs such as Fig.~\ref{fig:rge}(b).
%
%
%
%
%
%
%
%
%

\section{Anomalous Dimensions and Consistency Conditions}\label{sec:anom}

In this section, we generalize 't~Hooft's computation of anomalous dimensions and consistency conditions~\cite{tHooft:1973mfk} to an EFT, where the renormalization structure is non-linear. Consider a general EFT Lagrangian in $d=4-2\epsilon$ dimensions. Scalar and gauge fields have dimension $1-\epsilon$, and fermion fields have dimension $3/2-\epsilon$. An operator $O_i$ with a total number of fields $F_i$ has fractional dimension $-f_i \epsilon$, where $f_i \equiv F_i-2$. The Lagrangian is
\begin{align}
{\cal L} &= \sum_i  C_i^{(b)} O^{(b)}_i =  \sum_i \mu^{f_i \epsilon}  C_i O_i + \text{c.t.} \,,
\label{1.1}
\end{align}
where $C_i^{(b)}$ are the bare couplings, $O^{(b)}_i $ are the bare operators, $C_i$ are the renormalized couplings, and $\mu^{f_i \epsilon}$ are included to get the correct dimensions for ${\cal L}$ in $d$ dimensions. The coefficients
$C_i$ include couplings in the dimension-four part of the Lagrangian, as well as higher dimension coefficients in the EFT. In scalar theory, for example, the interaction term is $-\mu^{2\epsilon} \lambda \phi^4/4!$. Similarly, gauge and Yukawa couplings, which are cubic interactions, come with a factor $\mu^\epsilon$. The factor $f_i$ counts the factors of $4\pi$ associated with an operator in naive dimensional analysis~\cite{Manohar:1983md,Jenkins:2013sda,Gavela:2016bzc}. If an $L$ loop graph with insertions of $O_j$ produces an operator $O_i$, then there is a topological identity~\cite[(12)]{Gavela:2016bzc} which implies the naive dimensional analysis counting given in Eq.~\eqref{2.1},
\begin{align}
f_i + 2 L &= \sum_j f_j \,,
\label{1.2}
\end{align}
where $f_i+2$ are the number of external fields in $O_i$.

In the MS scheme, counterterms are poles in $\epsilon$. Since the EFT is non-linear the general renormalization structure is
\begin{align}
C^{(b)}_i \ \mu^{-f_i \epsilon}  =   C_i + \sum_{k=1}^\infty \frac{a^{(k)}_i \left(\{C_j\} \right) }{\epsilon^k} \,,
\label{1.3}
\end{align}
where $a^{(k)}_i \left(\{C_j\} \right)$ is a product of coefficients $C_{j_1} \ldots C_{j_n}$, and a given $C_j$ can appear multiple times. The usual gauge, Yukawa, and scalar self-couplings are included in the $\{C_j\} $. The counterterms satisfy  EFT power counting.
The counterterms $a^{(k)}_i $ have a loop expansion,
\begin{align}
a^{(k)}_i  &= \sum_L a^{(k,L)}_i \,.
\label{1.12}
\end{align}
We define the loop operator $\mathsf{L}$ by
\begin{align}
\mathsf{L} \, a^{(k)}_i  &\equiv \sum_L L   a^{(k,L)}_i \,.
\label{1.13}
\end{align}
Each vertex $C_{j}$ comes with a factor $\mu^{f_j \epsilon}$. 
The identity Eq.~\eqref{1.2} shows that the product of the $\mu^{f_j \epsilon}$ factors at the individual vertices gives the $\mu^{f_i \epsilon}$ factor needed for the operator $O_i$, as well as an additional  $\mu^{2L \epsilon}$ factor. Each loop integral has fractional dimension $-2\epsilon$, so the $L$-loop  integral gives
$(p^2)^{-L\epsilon}$, where $p$ is a generic mass or external momentum. The two contributions combine into the
dimensionless ratio $(\mu^2/p^2)^{L\epsilon}$, which gives logarithms of $\mu^2/p^2$ when expanded in $\epsilon$.

Differentiating Eq.~\eqref{1.3} w.r.t.\ $\log \mu$ gives
\begin{align}
- f_i \epsilon \left[ C_i + \sum_{k=1}^\infty \frac{a^{(k)}_i \left(\{C_j\} \right) }{\epsilon^k} \right] & =  \dot C_i  + \sum_j \sum_{k=1}^\infty \frac{1}{\epsilon^k} \frac{\partial a^{(k)}_i}{\partial C_j} \dot C_j
\label{1.4}
\end{align}
where $\dot C_j \equiv \mu\, \rd C_j/\rd \mu$. The order $\epsilon$ terms match on the two sides if
\begin{align}
\dot C_i &= -\epsilon  f_i  C_i + \gamma_i \,,
\label{1.5}
\end{align}
where $\gamma_i$ is independent of $\epsilon$.  Matching the order 1 term gives
\begin{align}
0 =  \gamma_i +  f_i a^{(1)}_i - \sum_j \frac{\partial a^{(1)}_i}{\partial C_j}\ f_j  C_j \,,
\label{1.6}
\end{align}
and matching the order $1/\epsilon^s$ term gives
\begin{align}
0 =  f_i  a^{(s+1)}_i -  \sum_j  \frac{\partial a^{(s+1)}_i }{\partial C_j} \ f_j  C_j  
+ \sum_j \frac{\partial a^{(s)}_i}{\partial C_j} \ \gamma_j \,, \qquad s \ge 1\,.
\label{1.7}
\end{align}
The identity Eq.~\eqref{1.2} implies
$a_i$ satisfies
\begin{align}
a^{(k)}_i (\{\lambda^{f_j} C_j\}) &= \lambda^{f_i+2 \mathsf{L}} a^{(k)}_i(\{C_j \})\,,
\label{1.8}
\end{align}
where $\lambda= \mu^\epsilon$ is a scale factor.
Differentiating Eq.~\eqref{1.8} w.r.t.\ $\lambda$ and setting $\lambda=1$ gives Euler's theorem on homogeneous functions
\begin{align}
\sum_j f_j C_j \frac{\partial a^{(k)}_i}{\partial C_j} &= (f_i+2\mathsf{L}) a^{(k)}_i \,.
\label{1.9}
\end{align}
Substituting this relation for $k=1$ into Eq.~\eqref{1.6} gives
\begin{align}
\gamma_i &= 2\mathsf{L} a^{(1)}_i \,.
\label{1.10}
\end{align}
Substitution into Eq.~\eqref{1.7} gives
\begin{align}
2 \mathsf{L} a^{(k+1)}_i &= \sum_j \gamma_j   \frac{\partial a^{(k)}_i }{\partial C_j}  = \mu \frac{{\rm d}}{{\rm d} \mu} a^{(k)}_i  ,
\label{1.11}
\end{align}
which is a consistency relation for the higher order poles.
The $1/\epsilon^2$ pole is given in terms of the $1/\epsilon$ poles by
\begin{align}
2 \mathsf{L} a^{(2)}_i &= \sum_j \gamma_j   \frac{\partial a^{(1)}_i }{\partial C_j} = 2 \sum_j \left( \mathsf{L} a^{(1)}_j \right)  \frac{\partial a^{(1)}_i }{\partial C_j} \,.
\label{1.11a}
\end{align}
Including a superscript $L$ for the loop order as in Eq.~\eqref{1.13}, one obtains the two-loop order results for Eq.~\eqref{1.10} and Eq.~\eqref{1.11a}
\begin{align}
\gamma^{(L=1)}_i &= 2 a^{(1,L=1)}_i \,, &
\gamma^{(L=2)}_i &= 4 a^{(1,L=2)}_i  \,, \nn
a_i^{(2,L=1)} & = 0\,, &
a^{(2,L=2)}_i &= \frac12 \sum_j a^{(1,L=1)}_j   \frac{\partial a^{(1,L=1)}_i }{\partial C_j} \,,
\label{1.14}
\end{align}
since the $1/\epsilon^2$ poles begin at two-loop order.

Equation~\eqref{1.10} gives the anomalous dimension in terms of the $1/\epsilon$ pole, and Eq.~\eqref{1.11} is the generalization of 't~Hooft's consistency condition for higher order poles. The anomalous dimension $\gamma_i\left(\{C_j\}\right)$ can be non-linear --- in SMEFT, the  dimension-eight evolution has terms proportional to the product of dimension-six coefficients, etc. The usual anomalous dimension $\gamma(g)$ in QCD also can be viewed as non-linear in $g$.
The usual forms of 't~Hooft's anomalous dimension and consistency equations in QCD involve terms such as
$g\, \partial a^{(k)}/\partial g$ which is simply $2 \mathsf{L} a^{(k)}$
since the $L$ loop contribution is order $g^{2L}$. Writing the results as Eq.~\eqref{1.10} and Eq.~\eqref{1.11} gives a simple form which generalizes to an arbitrary EFT with non-linear counterterms and RGE.

Consistency relations for the field anomalous dimension can be derived similarly. In paper II, we will see an example of an EFT with an infinite field anomalous dimension.\footnote{Some recent examples (which include the SM) are discussed in refs.~\cite{Bednyakov:2014pia,Herren:2017uxn,Herren:2021yur}. In the examples in paper II, the coupling constant anomalous dimensions are all finite.} To allow for this possibility, we take $\gamma_\phi$ to have an expansion in powers of $1/\epsilon$,
\begin{align}
\gamma_\phi &= \gamma_\phi^{(0)} + \sum_{k=1}^\infty \frac{1}{\epsilon^k} \gamma_\phi^{(k)} \,,
\end{align}
and $Z_\phi$ to have the expansion
\begin{align}
Z_\phi &= 1 + \sum_{k=1}^\infty \frac{z^{(k)}_\phi \left(\{C_j\} \right) }{\epsilon^k} \,.
\end{align}
Solving the equation $\dot Z_\phi = 2 Z_\phi \gamma_\phi$ gives
\begin{align}
- L   z_\phi^{(1)} &=  \gamma_\phi^{(0)}  \nn
-2 L   z_\phi^{(k+1)} + \frac{\partial z_\phi^{(k)}}{\partial C_j} \gamma_j  - 2   \gamma_\phi^{(0)} z_\phi^{(k)}  &= 
2 \left[ \gamma_\phi^{(k)}+ \sum _{r=1}^{k} \gamma_\phi^{(r)}z_\phi^{(k-r)} \right], \quad k \ge 1
\label{6.20}
\end{align}
The first equation in eq.~\eqref{6.20} determines the finite part of the field anomalous dimension $\gamma_\phi^{(0)}$. Requiring the l.h.s.\ of the second equation to vanish is the consistency condition for the field anomalous dimension. The failure of the consistency relations is given by the r.h.s. and comes from the divergent contributions to $\gamma_\phi$.

In the presence of evanescent operators, it is convenient to pick the subtraction scheme of Ref.~\cite{Dugan:1990df}, where insertions of evanescent operators do not contribute to $S$-matrix elements. This scheme involves subtracting additional finite contributions in Eq.~\eqref{1.3}. This changes the formul\ae\ for the renormalization group equations and consistency conditions~\cite{Dugan:1990df}. The generalization of these formul\ae\ to a general EFT will be presented in Ref.~\cite{Naterop:aa}.

\section{The $O(n)$ Model}\label{sec:on}

A detailed discussion of the application of our results to EFTs will be given in paper II, where we show how the geometric method greatly simplifies the computation.
Here we give a simple application of our results to the (renormalizable) $O(n)$ model, which illustrates the use of the formul\ae.  The Lagrangian is
\begin{align}
{\cal L} &= \frac12 (\partial_\mu \phi)\cdot(\partial^\mu \phi) -\Lambda -\frac12 m^2 (\phi \cdot\phi) - \frac14 \lambda (\phi\cdot \phi)^2  \,,
\end{align}
where $(\phi\cdot \phi) = (\phi^a \phi^a)$, $a=1,\ldots,n$.
The two-loop anomalous dimensions for this theory can be obtained from the general two-loop anomalous dimensions computed in Ref.~\cite{Machacek:1983tz,Machacek:1983fi,Machacek:1984zw}, which provides a check on our method. The shift $\phi \to \phi + \eta$ leads to Lagrangian terms quadratic, cubic, and quartic in $\eta$, respectively, given by
\begin{align}
\cL_{\eta^2} &= \frac12 (\partial_\mu \eta)\cdot(\partial^\mu \eta) -\frac12 m^2 (\eta \cdot\eta) -  \frac{\lambda}{2}  \left[    (\phi\cdot \phi) (\eta\cdot \eta)+2(\phi\cdot \eta)^2 \right]
\,,
\end{align}
\begin{align}
{\cal L}_{\eta^3} &=  -\lambda (\phi \cdot\eta) (\eta \cdot\eta)\,, & {\cal L}_{\eta^4}  &=- \frac14 \lambda (\eta\cdot \eta)^2\,.
\end{align}
Comparing with Eq.~\eqref{3.4a} and Eq.~\eqref{3.7} gives the quadratic coefficients
\begin{align}
N^\mu_{ab} &= 0, & X_{ab} &= -m^2 \delta_{ab} -\lambda \left[  \delta_{ab} (\phi \cdot\phi) + 2 \phi_a \phi_b \right]\,,
\end{align}
the cubic coefficients
\begin{align}
A_{abc} &= -\frac13 \lambda \left[ \delta_{ab} \phi_c + \delta_{bc} \phi_a + \delta_{ca} \phi_b \right]\,, & A^\mu_{a|bc} &= 0 \,,
\end{align}
and the quartic coefficients
\begin{align}
B_{abcd} &= -\frac1{12} \lambda \left[ \delta_{ab}\delta_{cd} + \delta_{ac} \delta_{bd} + \delta_{ad} \delta_{bc}  \right]\,, & B^\mu_{a|bcd} &= 0, & B^{\mu \nu}_{ab|cd} &=0\,.
\end{align}
$Y_{\mu\nu}=0$ since $N_\mu=0$. The one-loop counterterm from Eq.~\eqref{eq:oneloopct} is
\begin{align}
{\cal L}_{\text{c.t.}}^{(1)}  &=   \frac{1}{16 \pi^2 \epsilon}\left[- \frac14  n m^4 - \frac12 (n+2)\lambda m^2 (\phi \cdot\phi) - \frac14 (n+8) \lambda^2 (\phi\cdot \phi)^2 \right],
\end{align}
and the two-loop counterterm from Eq.~\eqref{eq:2loopBct} and Eq.~\eqref{eq:2loopAct} is
\begin{align}
{\cal L}_{\text{c.t.}}^{(2)}  &=  \frac{1}{(16\pi^2)^2} \biggl\{ -\frac{ n (n+2) \lambda  m^4 }{4 \epsilon ^2} -\frac{(n+2)\lambda ^2 }{4 \epsilon }(\partial\phi\cdot\partial\phi) -\left[
\frac{ (n+5)}{ \epsilon^2} - \frac{3}{\epsilon} \right] \frac {(n+2) \lambda ^2 m^2}{2} (\phi \cdot\phi) \nn
&  - \left[ \frac{(n+8)^2}{\epsilon^2} - \frac{ 2 (5n+22)}{\epsilon} \right] \frac14   \lambda ^3(\phi \cdot\phi )^2 \biggr\} \,.
\label{6.9}
\end{align}
From these counterterms, we get the renormalization group equations and anomalous dimensions up to two-loop order
\begin{align}
\mu\frac{\rd}{\rd \mu} \Lambda &=  \gamma_\Lambda \,,  &
\mu\frac{\rd}{\rd \mu} \log Z_\phi &= 2 \gamma_{\phi} \,, \nn
\mu\frac{\rd}{\rd \mu} m^2 &= \gamma_{m^2} \,, &
\mu\frac{\rd}{\rd \mu} \lambda &= \gamma_{\lambda} \,,
\end{align}
with
\begin{align}
\gamma_\Lambda &=  \frac{\lambda}{16\pi^2 } \frac12 n m^4 \,,\nn
\gamma_\phi &=   \frac{\lambda^2}{(16 \pi^2)^2} (n+2) \,, \nn
\gamma_{m^2} &= \frac{\lambda}{16\pi^2 } 2 (n+2) m^2 - \frac{\lambda^2}{(16\pi^2)^2} 10 (n+2) m^2 \,,   \nn
\gamma_\lambda &=  \frac{\lambda^2}{16\pi^2 }2  (n+8) - \frac{\lambda^3}{(16\pi^2)^2} 12 (3n+14)\,.
\end{align}
The anomalous dimensions given above only depend on the $1/\epsilon$ counterterms.
The $1/\epsilon^2$ two-loop counterterms are related to the $1/\epsilon$ one-loop counterterms by consistency conditions~\cite{tHooft:1973mfk}, which are satisfied. The anomalous dimensions agree with known results~\cite{Machacek:1983tz,Machacek:1983fi,Machacek:1984zw}, which can be conveniently cross-checked with \texttt{RGBeta} \cite{Thomsen:2021ncy}.

\section{Conclusions}\label{sec:conclusions}

We have presented the two-loop counterterms for scalar loops in a general EFT with interactions up to two derivatives in Eq.~\eqref{eq:2loopBct} and Eq.~\eqref{eq:2loopAct}. These counterterms formulae give the two-loop renormalization group equations. We found that factorizable graphs do not contribute to the RGE (with a possible finite subtraction in some cases). This is a general observation which extends to arbitrary loop order. The results were  applied to the renormalizable $O(n)$ model to obtain the two-loop RGE and to check the consistency conditions. We find agreement with the well-known results in the literature. 

The power of the formul\ae\ derived here is their application to effective field theories, which include higher dimension operators.  We present a geometric formalism in paper II, which allows for an efficient use of our results. We apply them to compute the two-loop renormalization for the $O(n)$ EFT, the scalar sector of SMEFT to dimension six, and chiral perturbation theory to order $p^6$.

The results presented here are for scalar loops. 't~Hooft showed in ref.~\cite{tHooft:1973bhk} that the scalar results could be used to also obtain the results for fermions and gauge bosons. The geometric method has been extended to fermion and gauge loops at one-loop order~\cite{Helset:2022pde,Helset:2022tlf,Finn:2020nvn,Gattus:2023gep,Assi:2023zid}.
The same procedure can be used at two-loops, but the algebra is more involved.

\subsection*{Acknowledgments}

We thank Tim Engel, Javier Fuentes-Mart\'in, Xiaochuan Lu, Chia-Hsien Shen,  Peter Stoffer and Anders Thomsen for helpful discussions.
This work is supported in part by the U.S.\ Department of Energy (DOE) under award numbers~DE-SC0009919. LN gratefully acknowledges financial support from the Swiss National Science Foundation (Project No.~PCEFP2\_194272 and mobility grant PCEFP2\_194272/2).

\begin{appendix}

\section{Extraction of UV divergences}
\label{appendix:uvextraction}

We describe how we obtain the UV divergences of nonfactorizing graphs, which, after subtraction of subdivergences, leads to the results in Table~\ref{tab:aact}. 
The method, introduced in \cite{Chetyrkin:1997fm}, rewrites the propagator using the identity
\begin{equation}
	\frac{1}{(k+p)^2-M^2} = \frac{1}{k^2-m^2} + \frac{M^2-p^2-2pk-m^2}{k^2-m^2}\frac{1}{(k+p)^2-M^2}
	\label{eq:tadpoledecomposition}
\end{equation}
where $M$ is a mass in the theory, $k$ a linear combination of loop momenta and $p$ a linear combination of external momenta. $m$ is an auxiliary mass introduced in intermediary steps of the calculation.  Eq.~\eqref{eq:tadpoledecomposition} is useful because the first term on the right-hand side is simpler and the second term has reduced degree of divergence. The decomposition eq.~\eqref{eq:tadpoledecomposition} allows one to express a two-loop integral in terms of simpler integrals which isolate the divergent parts, and more complicated expressions which are UV finite.

We will refer to denominators of the form $k^2-m^2$, which contain the auxiliary mass $m$ and \emph{no} external momenta as tadpole denominators. Eq.~\eqref{eq:tadpoledecomposition} converts denominators into tadpole denominators plus terms which fall off faster for large $k$.

At two loops, Feynman integrals can generally have double poles which are local (i.e. polynomial in the external scales), and single poles which can contain non-local functions of the external scales, such as $\log (p^2/\mu^2)$.
In \cite{Chetyrkin:1997fm} and \cite{Lang:2020nnl}, where a review of the method is given, the UV divergent parts are expressed purely in terms of tadpole integrals with  universal mass
\begin{equation}
	T(n_1,n_2,n_3) = \mu^{4\epsilon} \int_{k_1,k_2} \frac{1}{(k_1^2-m^2)^{n_1}(k_2^2-m^2)^{n_2}((k_1+k_2)^2-m^2)^{n_3}}
	\label{eq:tadpoleintegral}
\end{equation}
which generate non-local terms of the form $\log(m^2/\mu^2)$. In contrast to \cite{Chetyrkin:1997fm} and \cite{Lang:2020nnl}, we use a version of the method developed in \cite{Naterop:aa}, which  evaluates additionally terms non-local in the \textit{external} scales (such as $\log (p^2/\mu^2)$) explicitly, and checks the cancellation of these terms against one-loop diagrams with counterterm insertions. 

After discussing some fundamentals in Sec.~\ref{sec:appendix-fundamentals}, we apply our method to an explicit example in Sec.~\ref{sec:appendix-example} before describing the general algorithm in Sec.~\ref{sec:appendix-recursion}.

\subsection{Fundamentals}
\label{sec:appendix-fundamentals}
Let $k_1$ and $k_2$ be loop momenta and $p_k$ and $M_k$ be external momenta and particle masses. The auxiliary mass introduced by the algorithm is $m$. A generic two-loop tensor integral from Fig.~\ref{fig:two}(a) has the form 
\begin{equation}
    I = \mu^{4 \epsilon}  \int_{k_1,k_2}    \frac{k_1^{\mu_1}...k_1^{\mu_{r_1}}k_2^{\nu_1}...k_2^{\nu_{r_2}}}{ 
        \Big[\prod\limits_{i=1}^{b_1}\left((k_1+p_{1,i})^2-M_{1,i}^2\right)\Big]
        \Big[\prod\limits_{i=1}^{b_2}\left((k_2+p_{2,i})^2-M_{2,i}^2\right)\Big]
        \Big[\prod\limits_{i=1}^{b_3}\left((k_3+p_{3,i})^2-M_{3,i}^2\right)\Big]
        \label{a.3}
    }
\end{equation}
with $k_3 = k_1 + k_2$. There can be multiple propagators because the internal lines in Fig.~\ref{fig:two}(a) can have arbitrary insertions of quadratic vertices which can insert momentum. Let $C_i$ for $i \in \{1,2,3 \}$ denote the chain of propagators which only depends on the loop momentum $k_i$ corresponding to the three internal lines in Fig.~\ref{fig:two}(a). $I$ is UV finite if all subgraphs have negative superficial degree of divergence.\footnote{The graph itself is also treated as a subgraph.} Let $D_1$ be the degree of divergence
for $k_1 \to \infty$ with $k_2$ fixed, $D_2$ for $k_2 \to \infty$ with $k_1$ fixed, $D_3$ for $k_1 \to \infty$ with $k_1+k_2$ fixed, and $D_\text{G}$ for all $k_i \to \infty$ simultaneously.
$D_1$, $D_2$, $D_3$ are the degrees of divergence associated with the three subgraphs, and $D_\text{G}$ is the overall degree of divergence. 
 One therefore finds the following criterion for UV finiteness: $I$ is UV finite if $D_1$, $D_2$, $D_3$ and $D_\text{G}$ are all negative.
The values of $D_i$ are\footnote{The $r_i$ terms arise from the degree of divergence of the numerator of eq.~\eqref{a.3}. If the numerator is more complicated, one has to determine its behaviour in the various scaling limits. For example a numerator $k_1+k_2$ contributes $1$ to $D_1$, $D_2$, and $D_\text{G}$, but $0$ to $D_3$.}
\begin{equation}
    \begin{aligned}
        & D_1  = 4 + r_1 - 2 b_1 - 2 b_3 \\
        & D_2  = 4 + r_2 - 2 b_2 - 2 b_3 \\
        & D_3  = 4 + r_1 + r_2 - 2 b_1 - 2 b_2 \\
        & D_\text{G}  = 8 + r_1 + r_2 - 2b_1 - 2b_2 - 2b_3\,.
    \end{aligned}
    \label{eq:dod}
\end{equation}
Following \cite{Lang:2020nnl} we can think of eq.~\eqref{eq:tadpoledecomposition} as the action of an identity tadpole expansion operator
\begin{equation}
	\mathbbm{1}_i   = S_i + F_i
	\label{a.5}
\end{equation}
acting on a propagator of chain $C_i$ with $S_i$ and $F_i$ producing the first and second terms on the l.h.s.\ of eq.~\eqref{eq:tadpoledecomposition} respectively.

When acting recursively with tadpole expansion operators, it is possible that the recursion does not converge to just tadpole integrals and UV finite parts. We discuss an example in Sec.~\ref{sec:appendix-example}. We show in Sec.~\ref{sec:appendix-example} that this can be traced to stagnation in one of the degrees of divergence during repeated application of $F_i$. At this point we apply the \textit{disentangle identities}
\begin{equation}
	\frac{1}{(k_1+q_2)^2-M^2}  = \frac{1}{k_1^2-m^2} + \frac{M^2-q_2^2-2k_1q_2-m^2}{k_1^2-m^2}\frac{1}{(k_1+q_2)^2-M^2}
	\label{eq:disentangle1}
\end{equation}
\begin{equation}
	\frac{1}{(q_1+k_2)^2-M^2}  = \frac{1}{k_2^2-m^2} + \frac{M^2-q_1^2-2q_1k_2-m^2}{k_2^2-m^2}\frac{1}{(q_1+k_2)^2-M^2}
	\label{eq:disentangle2}
\end{equation}
where $q_1$ is $k_1$ plus a linear combination of external momenta, and similarly for $q_2$. These identities reduce the number of $k_1+k_2$ propagators in the UV divergent part.

Under application of these decompositions, the degrees of divergence (\ref{eq:dod}) generally change. For example, applying eq.~\eqref{eq:tadpoledecomposition} with $q = k_1$, the first term on the r.h.s.\ has unchanged degree of divergences compared to the l.h.s., but in the second term $D_1$ decreases by 1, $D_2$ remains unchanged, $D_3$ decreases by 1  and  $D_\text{G}$ decreases by one. We write this behaviour as $1^0 2^0 3^0 \text{G}^0 + 1^- 2^0 3^- \text{G}^-$. In table~\ref{tab:behaviour}, we summarize the behaviour of the decompositions eq.~\eqref{eq:tadpoledecomposition}, eq.~\eqref{eq:disentangle1}, eq.~\eqref{eq:disentangle2} used in the reduction algorithm.
%
%
\begin{table}
	\centering
	\begin{tabular}{l|l}
	decomposition            & behaviour                                      \\ \hline 
	\\[-1em]
	eq.~\eqref{eq:tadpoledecomposition} with $k=k_1$       & $1^0 2^0 3^0 \text{G}^0 + 1^- 2^0 3^- \text{G}^-$          \\
	eq.~\eqref{eq:tadpoledecomposition} with $k=k_2$       & $1^0 2^0 3^0 \text{G}^0 + 1^0 2^- 3^- \text{G}^-$          \\
	eq.~\eqref{eq:tadpoledecomposition} with $k=k_1 + k_2$ & $1^0 2^0 3^0 \text{G}^0 + 1^- 2^- 3^0 \text{G}^-$          \\
	eq.~\eqref{eq:disentangle1}                  & $1^0 2^{++} 3^{--} \text{G}^0 + 1^- 2^{++} 3^0 \text{G}^0$ \\
	eq.~\eqref{eq:disentangle2}                  & $1^{++} 2^0 3^{--} \text{G}^0 + 1^{++} 2^- 3^0 \text{G}^0$
	\end{tabular}
	\caption{\label{tab:behaviour} Behavior of degrees of divergence $D_i$ under application of the decomposition formul\ae~in the worst case scenario.}
\end{table}
%
%
Note that in eq.~\eqref{eq:disentangle1} and eq.~\eqref{eq:disentangle2}, the first term on the r.h.s.\ has no combined denominator. Therefore these terms lead towards factorized integrals. Note also that eq.~\eqref{eq:disentangle1} and eq.~\eqref{eq:disentangle2} increase some degrees of divergence. An important trick in the construction of the algorithm is to apply eq.~\eqref{eq:disentangle1} and eq.~\eqref{eq:disentangle2} only in situations where the \textit{maximum} degree of divergence does not change under eq.~\eqref{eq:disentangle1} and eq.~\eqref{eq:disentangle2}. In other words, before applying eq.~\eqref{eq:disentangle1} and eq.~\eqref{eq:disentangle2} we make sure that the $D_i$ which increases is sufficiently low so the maximum degree of divergence does not change. The algorithm discussed in Sec.~\ref{sec:appendix-recursion} is constructed in a way such that this is always the case. 

\subsection{An example}
\label{sec:appendix-example}
Before giving the general recipe we provide an example. Consider the integrand 
\begin{equation}
	I_0 = \frac{1}{\big[k_1^2-M_1^2\big][k_2^2-M_2^2\big][(k_1+k_2)^2-M_3^2\big]}.
\end{equation}
We would like to extract the UV divergent part of the integral $I$. The result is
\begin{equation}
	\begin{aligned}
		I = & \mu^{4 \epsilon} e^{2 \gamma \epsilon} \int \frac{d^D k_1}{\pi^{D/2}}\frac{d^D k_2}{\pi^{D/2}} I_0 =  -\frac{M_1^2 + M_2^2 + M_3^2}{2~\epsilon^2}
		\\ & + \frac{1}{\epsilon}\left[ -\frac{3}{2} \left( M_1^2 + M_2^2 + M_3^2 \right) \; + \; M_1^2 \log \frac{M_1^2}{\mu^2} +  M_2^2 \log \frac{M_2^2}{\mu^2} + M_3^2 \log \frac{M_3^2}{\mu^2}  \right] + \mathcal O(\epsilon^0). 
		\label{eq:exampleresult}
	\end{aligned}
\end{equation}
In our algorithm, the non-local terms $\log {M_i^2}/{\mu^2}$ come from factorized integrals generated by the algorithm. They cannot come from the tadpole integrals, since tadpole integrals never produce logs of external scales, only logarithms of the auxiliary mass.

$I$ has $D_1 = 0$, $D_2 = 0$, $D_3 = 0$ and $D_\text{G} = 2$. We now apply the tadpole decomposition eq.~\eqref{eq:tadpoledecomposition} to each chain, one by one,
\begin{equation}
	I_0 = \mathbbm{1}_1 \mathbbm{1}_2 \mathbbm{1}_3  I = (S_1 + F_1)(S_2 + F_2)(S_3 + F_3) I_0
\end{equation}
producing among other summands a term
\begin{equation}
	F_1 S_2 S_3 I_0 = \frac{M_1^2 - m^2}{\big[k_1^2-m^2\big]\big[k_1^2-M_1^2\big][k_2^2-m^2\big][(k_1+k_2)^2-m^2\big]}
\end{equation}
which has $D_1 = -2$, $D_2 = 0$, $D_3 = -2$, $D_\text{G} = 0$. Observe how $D_2$ did not change under application of $F_1$. Indeed, continued application of $(S_1 + F_1)$ yields a simple tadpole integral from $S_1$ but from $F_1$ a term 
\begin{equation}
	F_1 F_1 S_2 S_3 I_0 = \frac{(M_1^2 - m^2)^2}{\big[k_1^2-m^2\big]^2\big[k_1^2-M_1^2\big][k_2^2-m^2\big][(k_1+k_2)^2-m^2\big]}
	\label{eq:beforedisentangle}
\end{equation}
which has  $D_1 = -4$, $D_2 = 0$, $D_3 = -4$ and  $D_\text{G} = -2$. While $D_1$, $D_3$ and $D_\text{G}$ decrease under the action of $F_1$, $D_2$ remains zero. Continued application of $(S_1 + F_1)$ does not help, and neither does $(S_2 + F_2)$ or $(S_3 + F_3)$.  The reason is that the $k_2$ and $k_1+k_2$ denominators are already in tadpole form, so the identity eq.~\eqref{a.5} acts trivially, and leaves the integrand unchanged.

However, because $D_1$ is low enough, we can now afford to apply the disentangle identity eq.~\eqref{eq:disentangle2} even though it increases $D_1$ by two units.   After application of eq.~\eqref{eq:disentangle2} on eq.~\eqref{eq:beforedisentangle}, $D_1$  is still negative.   The second term is UV finite, and the first term is
\begin{equation}
	\frac{(M_1^2 - m^2)^2}{\big[k_1^2-m^2\big]^2\big[k_1^2-M_1^2\big]\big[k_2^2-m^2\big]^2},
\end{equation}
which is a product of two one-loop integrals. It evaluates to
\begin{equation}
	\frac{1}{\epsilon} \left[ m^2 + M_1^2 \left( -1 + \log \frac{M_1^2}{\mu^2} -  \log \frac{m^2}{\mu^2} \right)  \right] + \mathcal{O}(\epsilon^0),
\end{equation}
where the $\log$ involving $M_1$ matches the one in eq.~\eqref{eq:exampleresult}. The other non-local $M_2$ and $M_3$ pieces are similarly generated by disentangle identities in terms where $D_1$ or $D_3$ stagnates. This is a general feature of our method: non-local terms in external scales are generated by general (i.e. non-tadpole type) factorized integrals. In the end, the true divergence is recovered and all dependencies on the auxiliary mass cancel. 

\subsection{The recursion step}
\label{sec:appendix-recursion}

Our algorithm consists of recursive application of a reduction step, which we define in the following. Each reduction step takes as input a two-loop integral $I$ and outputs a sum of integrals $J_i$ resulting from the application of a certain decomposition on $I$. The recursion proceeds until we end up with tadpole integrals, factorized integrals and UV finite remainders. The goal is to pick the decomposition so that each resulting $J_i$ satisfies at least one of the criteria:
\begin{enumerate}[label=(\roman*)]
    \item $J_i$ is more tadpole-like than $I$ (it has less denominators containing external scales)
    \item $J_i$ is more disentangled than $I$ (it has less $k_1 + k_2$ denominators)
    \item $J_i$ is more finite than $I$ (it has reduced maximum degree of divergence)
\end{enumerate}
whilst \textit{not becoming worse} than $I$ in any of the other criteria. This can be achieved by the following recipe:
\begin{quote}
    { \textbf{Case 1 } If $\max{\{ D_1,D_2,D_3,D_\text{G} \} } = D_1$: }
    \begin{quote}
        \textbf{Case 1.1} \; If there is a non-tadpole $k_1$ or a non-tadpole $k_1 + k_2$ denominator, apply eq.~\eqref{eq:tadpoledecomposition} to it. \\
		\textbf{Case 1.2} \; Else if $D_2 < D_1 - 2$ apply eq.~\eqref{eq:disentangle1} on $((k_1+k_2)^2-m^2)^{-1}$ \\
        \textbf{Case 1.3} \; Else apply eq.~\eqref{eq:tadpoledecomposition} on a non-tadpole $k_2$ denominator. 
    \end{quote}
    { \textbf{Case 2 }  If $\max{\{ D_1,D_2,D_3,D_\text{G} \} } = D_2$: }
    \begin{quote}
        \textbf{Case 2.1} \; If there is a non-tadpole $k_2$ or a non-tadpole $k_3$ denominator, apply eq.~\eqref{eq:tadpoledecomposition} to it. \\
        \textbf{Case 2.2} \; Else if $D_1 < D_2 - 2$ apply eq~\eqref{eq:disentangle2} on $((k_1+k_2)^2-m^2)^{-1}$. \\
		\textbf{Case 2.3} \; Else apply eq.~\eqref{eq:tadpoledecomposition} on a non-tadpole $k_1$ denominator. 
    \end{quote}
    { \textbf{Case 3 }  If $\max{\{ D_1,D_2,D_3,D_\text{G} \} } = D_3$: }
    \begin{quote}
        \textbf{Case 3.1} \; If there is a non-tadpole $k_1$ or a non-tadpole $k_2$ denominator, apply eq.~\eqref{eq:tadpoledecomposition} to it. \\
        \textbf{Case 3.2} \; If $r_1 \leq r_2$ shift $k_1 \rightarrow -k_1 - k_2$ and go to \textbf{Case 2}, \\
		\textbf{Case 3.3} \;  else shift $k_2 \rightarrow -k_2-k_1 $ and go to \textbf{Case 1}
    \end{quote}
    { \textbf{Case 4 } If $\max{\{ D_1,D_2,D_3,D_\text{G} \} } = D_\text{G}$: }
    \begin{quote}
		Apply eq.~\eqref{eq:tadpoledecomposition} on any non-tadpole $k_1$, $k_2$ or $k_3$ denominator. 
    \end{quote}
\end{quote}
A few remarks: 
\begin{enumerate}
    \item We assumed that all $b_i>0$ because if any of the $b_i$ is non-positive we have a factorized integral. 
    \item In \textbf{Case 1.2}, we assumed the existence of a $k_1+k_2$ tadpole denominator. If this is not the case (and the conditions for \textbf{1.1} are not satisfied), there is no $k_1 + k_2$ denominator at all (neither tadpole type nor non-tadpole type), so we have a factorized integral, so nothing remains to be done. We have also assumed the presence of a non-tadpole $k_2$ denominator in \textbf{Case 1.3} if there is neither a non-tadpole $k_1$ nor a non-tadpole $k_1+k_2$ denominator. If this is not the case, we have a tadpole integral. The same reasoning applies to the assumptions made in \textbf{Case 2.2} and \textbf{Case 2.3}. 
    \item In \textbf{Case 4} we assumed the existence of at least one non-tadpole $k_1$, $k_2$ or $k_1+k_2$ denominator. If neither of these is present, the integral in question is a tadpole integral.
    \item In \textbf{Case 1.3} applying eq.~\eqref{eq:tadpoledecomposition} on a non-tadpole $k_2$ denominator results in a reduction of $D_2$ while the maximum degree of divergence is $D_1$. Therefore the terms generated in this case are not better than the original integral in terms of the conditions (i) -- (iii). However, since application of eq.~\eqref{eq:tadpoledecomposition} on a non-tadpole $k_2$ denominator does not introduce any new combined denominators, they are also not worse in (i) -- (iii). Any term fulfilling the criteria for \textbf{Case 1.3} will continue to do so until the action of \textbf{Case 1.3} has lowered $D_2$ enough, so that \textbf{Case 1.2} applies. Now \textbf{Case 1.2} makes each term more disentangled, meaning reducing the amount $k_1 + k_2$ denominators. Thus, in this case, it takes a finite number of steps to improve in (i) -- (iii). The same reasoning applies to \textbf{Case 2.3}.
\end{enumerate}

In summary, through application of the reduction step we have rewritten a general two-loop integral as a series of terms. Each such term is no worse in criteria (i) -- (iii) than the original integral. Each term is also better in at least one of those aspects, with the exception of the case discussed in remark 4 above, where improvement in at least one aspect takes not one, but a finite number of steps. After a finite number of recursive steps, we will thus end up with tadpoles, factorized integrals and UV finite remainders which can be discarded. By construction of the reduction step, the convergence is guaranteed.

In our calculation, we use \texttt{qgraf} \cite{Nogueira:1991ex} to generate diagrams and \texttt{Mathematica} to manipulate expressions. The library \texttt{\mbox{Package-X}}~\cite{Patel:2015tea,Patel:2016fam} is employed for evaluation of general one-loop integrals; for the two-loop tadpole integrals we implement our own \texttt{Mathematica} routines following \cite{Chetyrkin:1997fm}. In the evaluation of the Green functions $A^\mu A^\nu X Y$ and $A^\mu A^\nu Y Y$, we use \texttt{FORM} \cite{Vermaseren:2000nd,Ruijl:2017dtg} to deal with the large number of terms generated before the cancellation of subdivergences. 

\section{Evanescent Operators}
\label{app:evan}

In this appendix, we extend the results of Sec.~\ref{sec:anom} to include evanescent operators. We will use the scheme of ref.~\cite{Dugan:1990df}, in which evanescent operator insertions do not contribute to physical $S$-matrix elements. The scheme requires making additional finite subtractions beyond the usual subtraction of $1/\epsilon$ poles, which changes the anomalous dimensions and consistency conditions.

Divide the operators $O_i$ in the Lagrangian into physical operators $P_a$ and evanescent operators $E_\alpha$, 
\begin{align}
L &= C_a^{(P)} P_a + C_\alpha^{(E)}  E_\alpha \,.
\label{b.1}
\end{align}
The physical operators are linearly independent in $d=4$ dimensions, and the evanescent operators vanish in $d=4$.
At tree-level, the $S$-matrix is given by computing graphs with insertions of $P_a$ and $E_\alpha$. Any graph with one or more $E_\alpha$ insertions vanishes in $d=4$, so the $S$-matrix only depends on $C_a^{(P)}$, and we can drop $C_\alpha^{(E)}$. The amplitude from
$n$-loop graphs (including the counterterm insertions) is schematically
\begin{align}
\left(P + E \right) \left[\frac{1}{\epsilon^n} + \ldots + \frac{1}{\epsilon} + f \right] \prod C_a^{(P)} + \left(\epsilon P + E \right) \left[\frac{1}{\epsilon^n} + \ldots + \frac{1}{\epsilon} + f \right]  \prod C_a^{(P)}  \prod_{\alpha \ge 1} C_\alpha^{(E)} \,.
\label{b.2}
\end{align}
where $\alpha \ge 1 $ means the term has at least one evanescent coefficient.
The first term is the contribution of graphs with only physical operator insertions to a scattering amplitude. It can have singular terms in $\epsilon$ up to order $1/\epsilon^n$, and a finite piece denoted as $f$.
The graphs can generate physical amplitudes which are non-zero in $d=4$ as well as evanescent amplitudes. The second term is the contribution of graphs with at least one evanescent operator insertion to scattering amplitudes, which can again contribute to physical or evanescent amplitudes. The $1/\epsilon^k$ terms are \emph{local}, but the finite part $f$ can be non-local. The key observation is that if an insertion of $E$ generates $P$, it must vanish in $d=4$ and so is proportional to $\epsilon$, as shown explicitly in the second term. Multiple insertions of evanescent operators do not necessarily have higher powers of $\epsilon$. For example, if $\hat \mu$ is a fractional dimension index, $g_{\hat \mu \hat \mu}=-2\epsilon$ and $g_{\hat \mu \hat \nu} g_{\hat \mu \hat \nu}  = -2 \epsilon$ is still order $\epsilon$.

In the scheme of ref.~\cite{Dugan:1990df}, the counterterm is given by subtracting all the $1/\epsilon$ poles (which are local), \emph{including} the $\epsilon P$ piece in the second term, which gives a finite subtraction when multiplied by the $1/\epsilon$ divergence. The finite amplitude after adding the counterterm is
\begin{align}
\left(P + E \right)\, f \, \prod C_a^{(P)} + \left(\epsilon P + E \right) \, f \, \prod C_a^{(P)}  \prod_{j \ge 1} C_\alpha^{(E)} \,.
\label{b.3}
\end{align}
In $d=4$, the second term vanishes, so the $S$-matrix is again given only by $C_a^{(P)}$, and we can drop $C_\alpha^{(E)}$. This is the main advantage of the scheme of ref.~\cite{Dugan:1990df} --- evanescent coefficients do not contribute to physical scattering amplitudes in $d=4$.

The extra finite subtraction in terms involving $C_\alpha^{(E)}$ changes the formul\ae\ for the anomalous dimensions and consistency conditions given in Sec.~\ref{sec:anom}. Eq.~\eqref{1.3} is replaced by
\begin{align}
C^{(b)}_a \ \mu^{-f_a \epsilon}  =   C_a + a^{(0)}_a \left(\{C_j\} \right) + \sum_{k=1}^\infty \frac{a^{(k)}_a \left(\{C_j\} \right) }{\epsilon^k} \,, \nn 
C^{(b)}_\alpha \ \mu^{-f_\alpha \epsilon}  =   C_\alpha  + \sum_{k=1}^\infty \frac{a^{(k)}_\alpha \left(\{C_j\} \right) }{\epsilon^k} \,,
\label{b.4}
\end{align}
where $a^{(0)}_a$ is the additional finite subtraction which is only non-zero for physical operators.  The coefficients $a_\alpha^{(k)}\left(\{C_j\}\right)$ must contain at least one evanescent coefficient. Taking the derivative $\mu \frac{d}{d\mu}$ of Eq.~\eqref{1.3} yields
\begin{align}
-f_a \epsilon \left[ C_a + a^{(0)}_{a}  + \sum_{k\ge1} \frac{a^{(k)}_{a} }{\epsilon^k}  \right] &= \dot C_a + \sum_j \frac{\partial a^{(0)}_{a} }{\partial C_j} \dot C_j + 
\sum_{j} \sum_{k \ge 1} \frac{1}{\epsilon^k} \frac{\partial a^{(k)}_{a} }{\partial C_j} \dot C_j \nn
-f_\alpha \epsilon \left[ C_\alpha  + \sum_{k\ge1} \frac{a^{(k)}_{\alpha} }{\epsilon^k}  \right] &= \dot C_\alpha + 
\sum_{j} \sum_{k \ge 1} \frac{1}{\epsilon^k} \frac{\partial a^{(k)}_{\alpha} }{\partial C_j} \dot C_j 
\label{b.5}
\end{align}
where the sum on $j$ is over physical and evanescent indices, $j=\{b,\beta\}$.
Let $\dot C_i = -f_i  \epsilon C_i + \sigma_i \epsilon +  \gamma_i $ where $\sigma_i$ and $ \gamma_i$ do not depend on $\epsilon$. 
Let
\renewcommand{\arraycolsep}{0.25cm}
\renewcommand{\arraystretch}{1.25}
\begin{align}
M^{(k)}_{ij} &= \frac{\partial a^{(k)}_{i} }{\partial C_j}  & M^{(k)} &= \begin{bmatrix} M^{(k)}_{PP} & M^{(k)}_{PE} \\ M^{(k)}_{EP} & M^{(k)}_{EE} \end{bmatrix}
\label{b.15}
\end{align}
where the block diagonal form is in the space of physical and evanescent indices. For $k=0$,
\begin{align}
M^{(0)} &= \begin{bmatrix} M^{(0)}_{PP} & M^{(0)}_{PE} \\ 0 & 0 \end{bmatrix}, & \mathscr{M} = \mathbbm{1} + M^{(0)} ,
\label{b.17}
\end{align}
since $a^{(0)}_i$ is only non-zero if $i$ is a physical index. The matrix $\mathscr{M}$ is useful in deriving the renormalization group equations. Its inverse is
\begin{align}
\mathscr{M}^{-1}  =   \begin{bmatrix} (\mathbbm{1}+ M^{(0)}_{PP})^{-1}  & - (\mathbbm{1}+ M^{(0)}_{PP})^{-1} M^{(0)}_{PE} \\ 0 & \mathbbm{1} \end{bmatrix}
\label{b.16}
\end{align}
with vanishing $EP$ block.
Using the loop identity eq.~\eqref{2.1}, the order $\epsilon$ pieces of eq.~\eqref{b.5} give
\begin{align}
\begin{bmatrix} 2 \mathsf{L}  a^{(0)}_{a}  \\  0 \end{bmatrix} &= \mathscr{M}  \begin{bmatrix} \sigma_a  \\  \sigma_\alpha \end{bmatrix} ,
\label{b.9}
\end{align}
the order one pieces give
\begin{align}
\begin{bmatrix} 2 \mathsf{L}  a^{(1)}_{a}  \\    2 \mathsf{L}  a^{(1)}_{\alpha} \end{bmatrix} &=  \mathscr{M} \begin{bmatrix} \gamma_{a}  \\   \gamma_{\alpha} \end{bmatrix} 
+ M^{(1)} \begin{bmatrix} \sigma_a  \\  \sigma_\alpha \end{bmatrix} ,
\label{b.7}
\end{align}
and the $1/\epsilon^k$ pieces give
\begin{align}
\begin{bmatrix} 2 \mathsf{L}  a^{(k+1)}_{a}  \\    2 \mathsf{L}  a^{(k+1)}_{\alpha} \end{bmatrix} &= 
 M^{(k)}  \begin{bmatrix} \gamma_{a}  \\   \gamma_{\alpha} \end{bmatrix}  + M^{(k+1)} \begin{bmatrix} \sigma_a  \\  \sigma_\alpha \end{bmatrix} \,.
\label{b.11}
\end{align}
The solution of eq.~\eqref{b.9} is
\begin{align}
 \begin{bmatrix} \sigma_a  \\  \sigma_\alpha \end{bmatrix}  &= \mathscr{M}^{-1} \begin{bmatrix} 2 \mathsf{L}  a^{(0)}_{a}  \\  0 \end{bmatrix}
 =   \begin{bmatrix} (\mathbbm{1}+ M^{(0)}_{PP})^{-1}\,  (2 \mathsf{L}  a^{(0)}_{a})  \\ 0 \end{bmatrix}  \,.
\label{b.12}
\end{align}
The evanescent coefficients have $\dot C^{(E)}_\alpha = - f_\alpha C^{(E)}_\alpha $ with order $\epsilon$ contribution $-f_\alpha$, but the order $\epsilon$ term in the 
physical coefficient is modified, $\dot C^{(P)}_a = (- f_a C^{(P)}_a + \sigma_a ) \epsilon $ with $\sigma_a$ given by eq.~\eqref{b.12}.

The anomalous dimensions are obtained from eq.~\eqref{b.7},
\begin{align}
 \begin{bmatrix} \gamma_{a}  \\   \gamma_{\alpha} \end{bmatrix}  &=  \mathscr{M}^{-1} \begin{bmatrix} 2 \mathsf{L}  a^{(1)}_{a}  \\    2 \mathsf{L}  a^{(1)}_{\alpha} \end{bmatrix} 
-  \mathscr{M}^{-1}  M^{(1)} \mathscr{M}^{-1} \begin{bmatrix} 2 \mathsf{L}  a^{(0)}_{a}  \\  0 \end{bmatrix}
\label{b.21}
\end{align}
using eq.~\eqref{b.12}, and the consistency conditions from eq.~\eqref{b.11} are

\begin{align}
\begin{bmatrix} 2 \mathsf{L}  a^{(k+1)}_{a}  \\    2 \mathsf{L}  a^{(k+1)}_{\alpha} \end{bmatrix} &= 
 M^{(k)}  \begin{bmatrix} \gamma_{a}  \\   \gamma_{\alpha} \end{bmatrix}  + M^{(k+1)}   \begin{bmatrix} (\mathbbm{1}+ M^{(0)}_{PP})^{-1}\, ( 2 \mathsf{L}  a^{(0)}_{a} ) \\ 0 \end{bmatrix} \,.
\label{b.25}
\end{align}
Using eq.~\eqref{b.16}, the anomalous dimensions eq.~\eqref{b.21} become
\begin{align}
 \begin{bmatrix} \gamma_{a}  \\   \gamma_{\alpha} \end{bmatrix}  &= \begin{bmatrix} (\mathbbm{1}+ M^{(0)}_{PP})^{-1}  & - (\mathbbm{1}+ M^{(0)}_{PP})^{-1} M^{(0)}_{PE} \\ 0 & \mathbbm{1} \end{bmatrix} 
 \left\{  \begin{bmatrix} 2 \mathsf{L}  a^{(1)}_{a}  \\    2 \mathsf{L}  a^{(1)}_{\alpha} \end{bmatrix} \right.  -\left. \begin{bmatrix}
M^{(1)}_{PP} (\mathbbm{1}+ M^{(0)}_{PP})^{-1}  ( 2 \mathsf{L}  a^{(0)}_{a}  )  \\  M^{(1)}_{EP} (\mathbbm{1}+ M^{(0)}_{PP})^{-1}  (2 \mathsf{L}  a^{(0)}_{a})   \end{bmatrix} \right\}
\label{b.22}
\end{align}
%
which is far more complicated than eq.~\eqref{1.10}. The two-loop version of eq.~\eqref{b.22} was derived in ref.~\cite{Naterop:aa}.
Eq.~\eqref{b.22} reproduces the two-loop anomalous dimensions for the weak interactions given in ref.~\cite{Dugan:1990df}. In deriving the RGE for this case, note that the physical couplings involve not only the weak interaction coefficients but also the QCD gauge coupling $g$.
\end{appendix}

\bibliographystyle{JHEP}
\bibliography{refs.bib}

\end{document}